\documentclass[traditabstract]{aa}
\usepackage{txfonts}
\usepackage{graphicx}
\usepackage{array}
\usepackage{natbib}
\usepackage{multirow}
\usepackage{relsize}
\usepackage{amsbsy}

\begin{document}

\title{The extended epoch of galaxy formation: age dating 
of $\sim$3600 galaxies with $2<z<6.5$ in the VIMOS Ultra-Deep Survey
\thanks{Based on data obtained with the European 
	  Southern Observatory Very Large Telescope, Paranal, Chile, under Large
	  Program 185.A--0791.}}
\titlerunning{VUDS: The epoch of galaxy formation from age dating of galaxies at $2<z<6.5$}

 \author{R.~Thomas\inst{1,2}
\and O.~Le F\`evre\inst{1}
\and M.~Scodeggio\inst{3}
\and P. Cassata \inst{2}
\and B.~Garilli\inst{3}
\and V.~Le Brun\inst{1}
\and B.~C.~Lemaux \inst{1}
\and D.~Maccagni\inst{3}
\and J.~Pforr\inst{1}
\and L. A. M.~Tasca\inst{1}
\and G.~Zamorani \inst{4}
\and S.~Bardelli\inst{4}
\and N.P.~Hathi\inst{1}
\and L.~Tresse\inst{1}
\and E.~Zucca\inst{4}}
\institute{
Aix Marseille Universit\'e, CNRS, LAM (Laboratoire d'Astrophysique de Marseille) UMR 7326, 13388, Marseille, France
\and
Instituto de Fisica y Astronom\'ia, Facultad de Ciencias, Universidad de Valpara\'iso, Gran Breta$\rm{\tilde{n}}$a 1111, Playa Ancha, Valpara\'iso Chile
\and
INAF--IASF Milano, via Bassini 15, I--20133, Milano, Italy
\and
INAF--Osservatorio Astronomico di Bologna, via Ranzani, I--40127, Bologna, Italy\\
             \email{romain.thomas@uv.cl}
}
 \date{Received...; accepted...}
  \abstract{
In this paper we aim at improving constraints on the epoch of galaxy formation by measuring the ages of 3597 galaxies with reliable spectroscopic redshifts $2 \leq z\leq 6.5$ in the VIMOS Ultra Deep Survey (VUDS). We derive ages and other physical parameters from the simultaneous fitting with the GOSSIP+ software of observed UV rest-frame spectra and photometric data from the u-band up to 4.5 $\mu$m using composite stellar populations model spectra. We perform extensive simulations and conclude that at $z\geq2$ the joint analysis of spectroscopy and photometry combined with restricted age possibilities when taking into account the age of the Universe substantially reduces systematic uncertainties and degeneracies in the age derivation and that age measurements from this process are reliable.  
We find that galaxy ages range from very young with a few tens of million years to substantially evolved with ages up to 1.5 Gyr or more. This large age spread is similar for different age definitions including ages corresponding to the last major star formation event, stellar mass-weighted ages and ages corresponding to the time since the formation of 25\% of the stellar mass.
We derive the formation redshift $z_f$ from the measured ages and find galaxies that may have started forming stars as early as $z_f\sim15$. 
We produce the formation redshift function (FzF), the number of galaxies per unit volume formed at a redshift $z_f$, and compare the FzF in increasing observed redshift bins finding a remarkably constant 'universal' FzF. The FzF is parametrized with (1+z)$^\zeta$, with $\zeta \simeq 0.58\pm 0.06$, indicating a smooth increase of about 2 dex from the earliest redshifts $z\sim15$ to the lowest redshifts of our sample at $z\sim2$. Remarkably this observed increase in the number of forming galaxies is of the same order as the observed rise in the star formation rate density (SFRD). The ratio of the comoving SFRD with the FzF gives an average SFR per galaxy of $\sim7-17$M$_{\odot}/yr$ at $z\sim 4-6$, in agreement with the measured SFR for galaxies at these redshifts. 
From the smooth rise in the FzF we infer that the period of galaxy formation extends all the way from the highest possible formation redshifts that we can probe at $z\sim15$ down to redshifts $z\sim2$. This indicates that galaxy formation is a continuous process over cosmic time, with a higher number of galaxies forming at the peak in SFRD at $z\sim2$ than at earlier epochs. 
}

   \maketitle
%

\section{Introduction}
The time when galaxies formed remains poorly constrained. In the current $\Lambda$CDM hierarchical structure formation paradigm, dark matter (DM) halos form from the growth under gravity of early fluctuations in the matter density field, occasionally merging to form increasingly larger DM halos. Galaxies form as matter collapses in these deep potential wells, with gas cooling and fragmentation triggering star formation on galaxy scales \citep{White1978,Bromm2009}, as simulated in cosmological volumes (e.g. \citeauthor{Vogelsberger2014} \citeyear{Vogelsberger2014}). 

The epoch when the first galaxies are born still remains difficult to constrain observationally. This process is supposed to begin when Hydrogen in the Universe was being reionized, a time when photons emitted from young stars could hardly escape the neutral hydrogen still surrounding them. As young stars are born, they are supposed to rapidly ionize their immediate surroundings and create a bubble inside which the medium is fully ionized \citep{Bromm2009}. Growing bubbles then eventually overlap, fully ionizing the Universe.
This reionization process is a most fundamental step in galaxy evolution and its exact duration and end is the matter of considerable debate. 
The most recent results of cosmic microwave background (CMB) observations with the Planck satellite report that the reionization optical depth is $\tau=0.066\pm0.016$ leading to a redshift at which half of the Universe is reionized of $z_{re}$=8.8$^{+1.7}_{-1.4}$ \citep{Planck2015}, significantly later in cosmic time than was estimated from WMAP with $z_{re}=10.4$ \citep{Hinshaw2013}. 
However, the CMB results do not tell us when the main sources responsible for reionization formed,  and it does not contain information about galaxies forming after reionization is completed. 

Understanding when the populations of galaxies at different epochs in the Universe formed requires to perform a complete census of galaxies at the highest possible redshifts. 
The highest redshift galaxy candidates identified so far are at z$\simeq$10 \citep{Bouwens2014b,Bouwens2014a}, but beyond $z\sim8.7$ \citep{Zitrin2015} galaxies are identified solely based on photometric properties and spectroscopic confirmation is awaited for. Moreover, this confirmation rests mainly on the Lyman-${\alpha}$ emission line which is the only emission feature accessible with current facilities and such a line is rare at those redshift making these confirmations very difficult. 

Besides the direct identification of galaxies in the reionization epoch, another way to probe the formation of the first galaxies is to 
measure the ages of galaxies securely identified from spectroscopy at redshifts close to, but not necessarily into, the reionization epoch, and infer their redshift of formation. From the properties of observed galaxies with a formation redshift in the reionization era it is then possible to perform some form of galaxy archaeology in estimating what would have been the bulk properties of these galaxies during the reionization epoch.

Measuring ages of galaxies and associated limitations at increasingly higher redshifts was performed in
a number of studies. 
Using the SDSS sample \cite{Thomas2005} derived ages, total metallicities, and element ratios of 124
early-type galaxies in high and low density environments. They show that most star
formation activity in early-type galaxies happened between redshifts $\sim$3 and 5 in
high density and between redshifts 1 and 2 in low density environments. 
\cite{Cimatti2008} identified passive galaxies at 1.4$<$z$<$2 and from a stacked spectrum
infered ages of $\sim$1 Gyr, hence placing the formation redshift beyond z$\simeq$2.
\cite{Kaviraj2013} selected $\sim$330 spheroids at redshifts 1$<$z$<$3 and derived their age 
using spectral energy distribution (SED) fitting. They found that the star formation in these galaxies
likely peaked in the redshift range 2$<$z$<$5 with a median of z$\sim$3. 
At higher redshifts individual galaxy studies have reported ages placing their 
formation redshift beyond the end of reionization (e.g. \citeauthor{Ouchi2013} \citeyear{Ouchi2013}).
Beyond these exploratory studies the distribution of ages, and 
hence of formation redshifts, of the general population of galaxies at a given epoch is still unclear.
We do not know if galaxies formed coevally or over an extended period of cosmic time. 
The downsizing trends observed in galaxy studies \citep{Cowie1996,DeLucia2006} seem to indicate that the most massive galaxies ended their star formation first, and that this cessation of star formation progresses to lower masses as cosmic
time passes. A more extensive assessment of when galaxies formed their first stars would consolidate this picture. 

Comparing  the observed SED to stellar synthesis population models has long been recognized as a powerful method to measure galaxy ages, along with other important physical parameters like the stellar mass M$_{\star}$ and SFR as implemented in SED fitting codes, extensively described in \cite{Bolzonella2000,Thomas2005,Ilbert2006,Franzetti2008,Brammer2008}. 
While the process of measuring galaxy ages follows the same method as for M$_{\star}$ and SFR measurements, measuring the age of galaxies is traditionally considered as a more uncertain parameter. However, a study of galaxy ages at $z\gtrsim2$ would allow the use of the natural age upper limit given by the age of the Universe. Thus, one may expect that getting to very high redshifts considerably helps to limit the degeneracies plaguing galaxy age investigations at $z\lesssim2.$

Here we present a study of the ages of an unprecedented sample of 3597 galaxies selected from the VIMOS Ultra Deep Survey (VUDS)
with spectroscopic redshifts 2$<$z$<$6.5. These galaxies are being observed at early cosmic epochs thus they must have formed at even earlier
times. We may therefore identify galaxies which have formed most of their stars when the Universe was being reionized.
The age of galaxies in our sample is derived using a novel technique applying SED fitting to the combination of spectra observed with VIMOS on the VLT with broad-band photometry. We use extensive simulations to estimate uncertainties related to this process. From the ages and star formation histories we study the distribution of ages, derive the formation redshift distribution, and discuss consequences on the epoch of galaxy formation.

The paper is organized as follows. In Section \ref{data} we summarize the VUDS sample.
We present the technique developed to fit simultaneously observed spectra and photometric data
with the GOSSIP+ software and discuss the benefit of this approach in Section \ref{GOSSIP+}.
We present several age definitions in Section \ref{sec_age_def}.
The reliability of galaxy age estimates and tests for possible degeneracies 
at z$>2$ are discussed in Section \ref{simul}. 
The age distribution of galaxies with $2\leq z \leq 6.5$
is presented in Section \ref{age} and we derive and discuss the distribution in formation redshifts in Section \ref{sec_fzf}.
Results are discussed in Section \ref{discuss} before concluding.

We use a cosmology with $H_0=70~km~s^{-1}~Mpc^{-1}$, 
$\Omega_{0,\Lambda}=0.7$ and $\Omega_{0,m}=0.3$. 
All magnitudes are given in the AB system.

\section{Data: the VUDS survey and associated photometry}
\label{data}

We draw our sample from the VIMOS Ultra Deep Survey (VUDS), described in detail in \cite{lefevre15}; a short summary is provided below. 

VUDS was designed to study the evolution of galaxies in the very early Universe. The main selection criterion is based on photometric redshifts with $z_{phot} + 1\sigma \geq 2.4$, and $i_{AB} \leq 25$. The secondary peak of the photometric redshift probability function is also taken into account if it satisfies $z_{phot} + 1\sigma \geq 2.4$. In addition, objects in the Lyman break area of colour-colour plots (e.g., \textit{g}-\textit{r}, \textit{r}-\textit{i}) are also selected in case they are not already picked-up by the photometric redshift selection; this sample represents about 10\% of the whole target sample. 

Observations are conducted on the ESO-VLT with the VIMOS spectrograph \citep{OLF2003}. Spectra are observed for $\sim$14h in each of the LRBLUE and LRRED grisms with a spectral resolution $R\sim230$ and cover a wavelength range from $3650$\AA~to $9350$\AA. 

Data are reduced with the VIPGI software \citep{Scodeggio:05} and spectroscopic redshifts are measured with the EZ software \citep{Garilli2010}. EZ is based on the cross correlation of observed spectra with reference spectra, followed by visual inspection of each spectrum by two independent observers who then compare their measurements to set the final 'best' redshift measurement. A reliability flag is assigned to each redshift measurement, and gives the probability level for the redshift to be right, as described in \cite{lefevre15}. 

All VUDS galaxies are matched to the deep photometric catalogues available in each of the 3 VUDS fields: COSMOS, ECDFS and VVDS-02h \citep{lefevre15}. In the following we use all the broad-band optical and near-infrared photometric data available in these fields. In the COSMOS field this includes u band data from the CFHT Legacy Survey, griz bands from Subaru \citep{taniguchi07}, YJHK photometry from the UltraVista survey reaching $K_{AB}=24.8$ at $5 \sigma$ \citep{McCracken2012}, as well as 3.6 and 4.5 micron Spitzer data from SPLASH (\citealt{Steinhardt2014}, Laigle et al in prep). The ECDFS has deep UBVRI imaging down to R$_{AB} = 25.3$ (5$\sigma$, \citealt{Cardamone:10}, and reference therein), WFC3 near-IR imaging that reaches as deep as $H_{AB}=27.3-27.6$ in the CANDELS survey (\citealt{Grogin2011},\citealt{Koekemoer11}), and Spitzer-warm 3.6 and 4.5 $\mu$m imaging data down to $AB=23.1$ from the warm Spitzer survey SERVS \citep{Mauduit12}. In the VVDS-02h field $u', g, r, i$ observations are available from the CFHTLS survey reaching $i_{AB}=25.44$ at 50\% completeness in the latest DR7 \citep{Cuillandre12}. Deep infrared imaging was obtained with WIRCAM at CFHT in YJHK bands down to $Ks_{AB}=24.8$ also at 50\% completeness \citep{Bielby:12}.

Spectra are calibrated in flux in a two step process. First a flux calibration is performed using the standard ESO calibration observations of spectrophotometric standard stars using the same VIMOS set-up. A further iteration is applied on the spectra to correct for two well-known, wavelength-dependent additional effects: atmospheric transmission that depends on the air-mass of the observations and atmospheric refraction that drives a fraction of an object's light out of the slit. Assuming that the light of the galaxy entering the slit is representative of the host object these corrections are applied on an object-by-object basis, as described in  \cite{Thomas14}. The flux scale of each spectrum is normalized to the observed i-band photometric flux derived from the total magnitude estimated using the mag-auto parameter from SExtractor \citep{Bertin:96}. This corrects for slit losses due to the 1 arcsecond slit width used for the observations, which is generally smaller than observed object sizes. At the end of the calibration process we compare broad-band magnitudes derived from the imaging data and broad-band magnitudes computed by integrating the flux of the spectrum given the same filter transmission curves. We find that the differences between the photometric and spectroscopic flux in the broad-bands u filters is 0.02$\pm$0.3, while the difference in i-band is null, by definition. 

The sample of VUDS galaxies that we are using in this paper is presented in section \ref{galsample}.


\section{Combining spectra and photometry to measure ages and other physical parameters of galaxies: the GOSSIP+ tool}
\label{GOSSIP+}

SED fitting has reached a maturity level such that SFR and M$_{\star}$ are extensively used in the literature \citep[e.g.][]{Ilbert2006,Ilbert09,Ilbert2013}. 
From the stellar population models one can also recover the age(s) of the stellar population(s) assembled in the model. 

At $z<2$, age estimations are prone to degeneracies and ages are used only when derived from extensive analysis combining SED fitting with some specific spectral indicators able to break the degeneracies (see e.g. \citeauthor{Thomas2005} \citeyear{Thomas2005}). Using photometry alone the degeneracies between age, metallicity, dust extinction and star formation history appear quite significant (e.g. \citealt{Wuyts09}, \citealt{Pforr12}). Therefore age measurements derived from SED fitting alone are often taken with skepticism. As a consequence, while the computation of M$_{\star}$ and SFR is well documented and those parameters are extensively used in the literature, ages of galaxies are only rarely the subject of detailed analysis. 

In this Section we present an improved method to determine ages based on the joint fitting of observed spectra and multi-band photometry. 

\subsection{Method}

The classical approach to measure photometric redshifts together with physical properties of galaxies (M$_{\star}$, SFR, dust extinction, age, metallicity) is to perform SED fitting of broad-band photometric measurements with stellar population synthesis models (e.g. \citeauthor{Bolzonella2000} \citeyear{Bolzonella2000}, \citeauthor{Ilbert2006} \citeyear{Ilbert2006}, \citeauthor{Brammer2008} \citeyear{Brammer2008}, \citeauthor{Maraston2010} \citeyear{Maraston2010}, \citealt{Pforr13}). This process makes use of a small number of data points, ideally a dozen of broad-band measurements ranging from the u-band to the K-band or beyond (see e.g. \citeauthor{Ilbert2013} \citeyear{Ilbert2013}), nevertheless providing remarkable constraints on main parameters like stellar mass.

When spectroscopy is available in addition to photometric data the information on the SED is significantly increased. The VIMOS spectra of VUDS are sampled by about 1000 spectral data points. The wavelength coverage ranges from $3650$\AA~to $9350$\AA. At $z>2$, this is sampling the UV rest frame and has the potential to improve constrains on recent star formation and dust extinction. The strong constraint provided on the strength and slope of the UV continuum \citep{Hathi15} and IGM transmission \citep{Thomas14}, UV rest-frame spectroscopy can help in reducing degeneracies that are typical when using the more coarse wavelength sampling of broad-band photometry.

Based on these considerations, we have developed a methodology able to benefit from both a set of photometric data points covering a large wavelength base but with a poor wavelength sampling, and spectroscopic data covering a smaller wavelength range but with high sampling. This method is applied to the VUDS data but is of general interest  when both spectroscopic and photometric datasets are available and complement each other in wavelength.  

\subsection{Joint spectroscopic and photometric fitting with stellar population synthesis models with GOSSIP+}

GOSSIP (Galaxy Observed-Simulated SED Interactive Program) is a tool built to perform the fitting of both spectroscopy and photometry with stellar population models.  It was initially developed \citep{Franzetti2008} to be used in the framework of the VVDS \citep{LeFevre2005,LeFevre2013} and zCOSMOS \citep{Lilly2007} surveys. The novelty of this software is the fact that it allows, for a given galaxy with a known redshift, for the combination of photometric measurements in different bands with spectroscopic data and performs a $\chi^{2}$ minimization fitting against a set of synthetic galaxy spectra based on emission from stellar populations. The result of the fit is used to estimate several parameters including SFR, age, and M$_{\star}$ of the observed galaxy through the Probability Distribution Function (PDF) of each parameter. GOSSIP can be used with different synthetic templates, which can be computed from reference models like  \cite{BC03} (hereafter BC03) as well as Maraston models (\citealt{Maraston2005}, \citealt{Maraston2011}, hereafter M05 and M11, respectively). Other user-defined models can also be implemented in GOSSIP.

We developed a new version of the software, called GOSSIP+, in order to add new functionalities and to modify pre-existing functions. We describe the main modifications that have been implemented in the next subsections. 

\subsubsection{Use of all spectral points}
In the original version of the software \citep{Franzetti2008}, the spectroscopic fitting was performed through a rebinning of the spectrum in larger wavelength bins. The new version of the software, GOSSIP+, makes now use of all the observed spectroscopic points directly when performing the fit. The only constraint is that reference models need to have a spectral resolution equal or better to the spectral resolution of the observed data otherwise the data may need to be resampled to the lower resolution of the models. 

\subsubsection{Template Builder}
A template builder is implemented. This allows to compute composite stellar population (CSP) models from a library of simple stellar populations (SSP) provided by BC03, M05 or M11. The SSPs are defined by two main parameters: the metallicity and the initial mass function (IMF). BC03 allows to explore metallicities from Z=0.0004 to Z=0.05 while M05 metallicities range from Z=0.001 to Z=0.04. The IMF can be either \cite{Salpeter1955}, \cite{Chabrier2003} or \cite{Kroupa2001}. From those SSPs, GOSSIP+ creates a CSP by applying a star formation history (SFH). Various SFH can be used, including exponentially declining SFH (SFH $\propto \tau^{-1} exp(-t/\tau)$) or delayed $\tau$ models SFH (SFH $\propto t \times \tau^{-2} exp(-t/\tau)$). The extinction by dust in a galaxy is then applied using a Calzetti law \citep{Calzetti2000}. The age of each extracted template is defined by the user between a few hundred thousands years up to 20 Gyr\footnote{BC03 and M05/M11 templates are intrinsically defined on a age grid of 221 ages.}


\subsubsection{IGM prescription}
Another new feature implemented in GOSSIP+ is the treatment of the Inter-Galactic Medium (IGM). The UV rest-frame spectra of distant galaxies combine intrinsic emission with absorption produced by intergalactic gas clouds present along the line of sight. The transmission of the IGM is an important property to be taken into account when analysing the SED of galaxies at z$>$1.5 \citep{Madau1995}, and models are systematically used when searching for and analysing distant galaxy spectra. Below this redshift the transmission of the IGM reaches $\sim$100\% because the intervening clouds along the line of sight are not numerous enough and the IGM contribution can be neglected, but above this redshift the IGM transmission becomes substantial and severely modifies the spectral shape blue-ward of the Lyman-$\alpha$ line. 

Several models have been proposed to estimate the transmission of the intergalactic medium \citep[e.g.][]{Madau1995,Meiksin2006}. Those models provide, for a given redshift, one single average IGM transmission as a function of wavelength, and this mean value is used by SED fitting algorithms. 
However, using a mean value for the IGM transmission is a simplifying assumption for ensemble averages of large samples of galaxies and does not necessarily translate correctly to the measurements of individual galaxies as not all lines of sight on the sky are populated with the same density or distribution of HI clouds.
In \citet{Thomas14} we show that the IGM transmission towards distant galaxies presents a large range of values at any given redshift. While we find a mean transmission as predicted by \cite{Meiksin2006} and in agreement with quasar studies (e.g., Beker et al, 2015), we find that the dispersion is large at any given redshift $2.5<z<5.5$. 

To take into account the range of possible IGM transmission, we consider the IGM as a free parameter in GOSSIP+ adding to the mean IGM transmission curve from \cite{Meiksin2006} six additional IGM transmission curves that span the observed range of transmission at any given redshift \citep{Thomas14}. With this prescription the IGM transmission at 1100\AA~ can vary from 20\% to 100\% at z=3.0 and from 5\% to 50\% at z=5.0.

\subsubsection{Emission lines}
It is now  widely documented (e.g. \citealt{debarros14}) that emission lines in star-forming galaxies (SFGs), the dominant population at high redshifts, must be taken into account in the model template spectra used in the cross-correlation with the observed SED. Strong emission lines present in SFGs such as H$\alpha$, the H$\beta$ and [OIII]4959/5007\AA ~doublet, [OII]3737\AA, and Ly$\alpha$, may change observed broad-band magnitudes by several tenths of magnitudes. Physical parameters derived from SED fitting may therefore be wrongly estimated if model spectra used in the SED fitting do not include these emission lines \citep[e.g.][]{Ilbert09,debarros14}.  

We have therefore implemented a treatment of emission lines in GOSSIP+. For the purpose of this analysis we opted to model emission lines using standard photo-ionization case-B recombination.
Six emission lines are implemented:  $Ly\alpha$, [OII], [OIII$_{a,b}$], [H$\beta$] and [H$\alpha$].
The procedure implemented in GOSSIP+ works as follows. The rescaling factor between the synthetic model and the observation gives an access to the SFR of the template. This SFR is then converted into [OII] Luminosity with the Kennicut law \citep{Kennicutt98}. The emission lines are then estimated using the following line ratios: [OIII/OII] = 0.36, [H$\beta$/OII]=0.61, [H$\alpha$/OII]=1.77 and [Ly$\alpha$/OII]=2, similar to what is used in the LePhare software \citep{Ilbert09}. As the Ly$\alpha$ line can be either in absorption or in emission this receives a particular treatment. A rough measurement of the equivalent width of the Ly$\alpha$ line is computed directly on the observed spectrum, if EW(Ly$\alpha$)<-10\AA~ (here the minus sign means emission) the emission line is created following the above ratios. If EW(Ly$\alpha$)$>-10$\AA~the emission line is not taken into account as it is not expected to affect broad-band magnitude measurements. One may argue that taking into account the Ly$\alpha$ line may not be done correctly since it does not linearly trace the SFR. To test for this effect we fitted a sample of strong Ly$\alpha$ emitters (EW(Ly$\alpha$)$<-25$\AA), with and without taking into account the Ly$\alpha$ line. The results for M$_{\star}$ and SFR are not significantly modified while age measurements are not affected at all.

\subsection{$\chi^{2}$ and PDF computation}

The fit of a spectrum or photometric data follows the same recipe as standard SED-fitting codes \citep{Bolzonella2000,Ilbert2006,Brammer2008,Maraston2010}. 
To fit the spectra and photometry of VUDS galaxies we first create libraries of template models from BC03 and M05. Then for each observed galaxy, we perform a $\chi^{2}$ minimisation over the entire template library. For a given galaxy and a given template the $\chi^{2}$ is computed with:

\begin{equation}
\chi^{2}_{i}=\sum^{N}_{i=1}\frac{F_{obs,i}-A\times F_{temp,i}}{\sigma_i},
\label{chi2form}
\end{equation}

where $F_{obs,i}$, $F_{temp,i}$, $\sigma_i$ and $A$ are the observed flux (or magnitude), the synthetic flux density (or magnitude) from the template, the observed error, and the normalisation factor applied to the template, respectively. The normalisation factor is computed from the comparison of the observed broad-band  photometry and the photometry computed on the galaxy template.
The reduced $\chi^2$, $\chi^2_{\nu}$, is then computed as in \cite{Salim2007} 
by considering that the number of parameters that is fitted is not linked to the number of physical parameters (i.e, properties) of the galaxies. The parameter that is fitted is then $A$. Consequently, the number of degrees of freedom is given by $N-1$ where $N$ is the number of data points.\\
For a given observation GOSSIP+ computes the $\chi^{2}$ of all the templates. The set of $\chi^{2}$ values are then used to create the probability distribution function (PDF). To build this PDF we compute for each template the probability:
\begin{equation}
P_{i}=\exp(\frac{-\chi^{2}_{i}}{2})
\label{proba_G}
\end{equation}
We use the median of the normalised PDF to estimate the parameter value. The errors on each parameter are taken from effective $\pm1 \sigma$ values of the PDF derived from the values including 68\% of the PDF of each parameter centred on the median value of the PDF.\\

When GOSSIP+ combines both a photometry dataset and a spectrum it has to take into account that the weight of both datasets are not the same. The spectroscopy contains $\sim$1000 points for VUDS-like spectra while the photometry is made of $\sim$10 points. Then if both observations are not weighted one photometric point will have the same importance as a spectroscopic point, and the spectrum will dominate the fit. We decided to give the same weight to both photometry and spectroscopy. Therefore, the combination of the full spectrum and the full photometry in one single fit is performed with the combined $\chi^{2}_{comb}$ defined as:
\begin{equation}
\chi^{2}_{comb}=\chi^{2}_{\nu}(\mathrm{phot}) + \chi^{2}_{\nu}(\mathrm{Spec}),
\end{equation}
where, $\chi^{2}_{\nu}(\mathrm{phot})$ and $\chi^{2}_{\nu}(\mathrm{Spec})$ are the reduced $\chi^{2}$ of the photometric and spectral fits, respectively. 

To compute the PDF of the combined fit we therefore choose to give the same weight to the large ensemble of points of the spectrum and to the smaller number of photometric points. This choice rests on the spectral coverage and SNR of the VUDS spectra on the one hand and on the broad wavelength coverage of the photometry on the other hand. Different weighting schemes may be applied for different datasets, giving more or less relative weight to the photometry and spectroscopy depending e.g. on the wavelength range  and spectral resolution of the spectra or the wavelength coverage and depth of the photometric data. While a complete statistical treatment of the relative weights between spectroscopy and photometry remains to be defined, we emphasize that changes of our weighting scheme by a factor of two do not significantly modify the results presented in this paper.

\subsection{Selection of reliable fits}
\label{galsample}

In addition to selecting VUDS galaxies with spectroscopic reliability level of 50, 75, 95, 100, and 80\% (flags 1,2,3,4, and 9, and the like, respectively, see \citealt{lefevre15}) we perform a selection based on the quality of the fit as described below.

Our fitting method is based on the ability to combine both the photometry and spectroscopy. The agreement between spectroscopy and photometry is therefore crucial to produce a good fit. As it can happen that there is a mismatch between these two datasets affecting the quality of fit we implement a visual inspection of each fit to define its quality. 

Four fit quality flags are used, as described in Table \ref{fitflag}. 
\begin{table}[h!]
\caption{Fit quality flag used in our study to select the best galaxy sample.}
\label{fitflag}    
\centering                       
\begin{tabular}{c c  }       
\hline\hline                
Fit Flag & Meaning  \\   
\hline                       
0 & Bad fit  \\ 
1 &  Poor fit  \\ 
2 & Good fit \\
3 & Excellent fit \\ 
\hline                                  
\end{tabular}
\end{table}
The two weakest flags, 0 and 1, correspond to bad and poor fits, respectively. These flags indicate a significant departure of the fit compared to either of the photometric or spectroscopic data points, or both. A bad fit can also be the result of an incorrect spectroscopic redshift assignment, becoming evident when using a broad wavelength range and the fit. Galaxies with fit quality flags 2 and 3 correspond to good to excellent fits, and these constitute the working sample for our study.

\begin{figure}[h!]
\centering
\includegraphics[width=8cm,height=8cm]{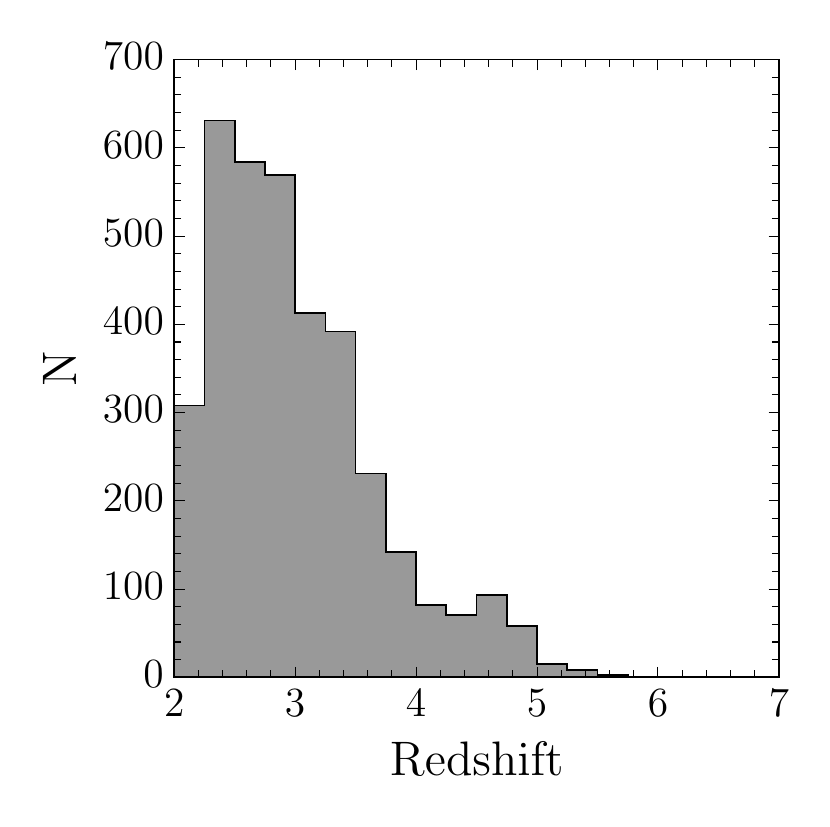}
\caption{Redshift distribution of the 3597 VUDS galaxies at $z>2$ used in this study.}
\label{Epoch_OBS}
\end{figure}

We verified that the visual selection mainly selects the fits with the smallest $\chi^{2}_{\mathrm{comb}} $, for which both the photometry and spectroscopy are well reproduced by a galaxy model (see Section \ref{fitproc} for examples). The fraction of selected galaxies with $\chi^{2}_{\mathrm{comb}} >10$, corresponding to increasingly bad fits, is negligible. As expected the visual classification follows the redshift reliability classification of the survey \citep[spectroscopic flag, see ][]{lefevre15}. 
The visual classification therefore retains galaxies with both a reliable spectrum and SED fit and a reliable spectroscopic redshift.   
After this selection we selected $\sim 67$\% of the total $z\geq 2$ VUDS galaxies. 
Our working catalogue is composed of 3597 VUDS galaxies with redshift between $z=2$ and $z=6.5$ that have a Fit Flag of 2 or 3. The redshift distribution of our sample is presented in Figure \ref{Epoch_OBS}.
This sample is a representative sample of the general star--forming population in the VUDS survey with M$_{\star}$ and SFR distributions similar to the full
sample.




\section{The ages of galaxies: definitions}
\label{sec_age_def}

The age of a galaxy is a parameter that can be defined in many different and complementary ways. One may want to find the time when the first stars where formed, or the time in a galaxy's life when a fraction of its stellar mass was assembled, as defined below.  

\begin{figure}[h!]
\centering
\includegraphics[width=8cm,height=8cm]{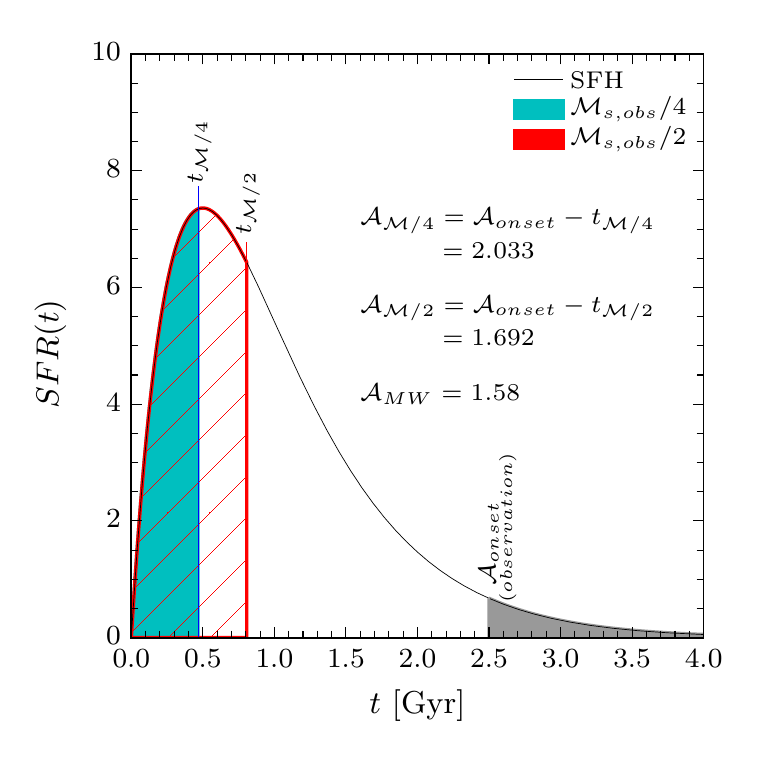}
\caption{Different Age definitions and their relation to stellar mass on an example. The grey line represents the SFH of the galaxy while the grey shaded area is in the \textit{future} of the observation (the '0' of the x-axis represents the formation of the galaxy). The blue and red areas represent one quarter and half of the stellar mass, respectively. Finalle, $t_{M/4}$ and $t_{M/2}$ show the time since the onset of the star formation and the time to build up 25\% and 50\% of the stellar mass, respectively. In this example the galaxy is observed 2.5 Gyr after the start of star formation, $\mathcal{A}_{onset}=2.5$ Gyr, and the stellar mass is $10^{10}\mathcal{M}_{\odot}$. This induces $\mathcal{A}_{M/4}=2.033$ Gyr, $\mathcal{A}_{M/2}=1.692$ Gyr and $\mathcal{A}_{MW}=1.58$ Gyr.  }
\label{Age-def-schema}
\end{figure}

We refer the reader to Figure \ref{Age-def-schema} for a visual interpretation of the three definitions listed below. 

We define $\mathcal{A}_{onset}$ as the age corresponding to the onset of star formation, i.e. the time since the beginning of the star formation history (SFH) that corresponds to the beginning of the stellar mass assembly of the galaxy. This definition corresponds to the output of the standard stellar population models with smooth SFH used to perform SED-fitting.  $\mathcal{A}_{onset}$ is related to the observed M$_{\star}$ by the following relation (see Fig.\ref{Age-def-schema}):

\begin{equation}
\mathcal{M}_{\star,obs}=\int^{\mathcal{A}_{onset}}_{0}SFH(t)dt,
\end{equation}
where $\mathcal{M}_{\star,obs}$ and $SFH(t)$ are the observed M$_{\star}$ and SFH of the galaxy, respectively.

One may argue that defining the \textit{starting point} of the life of a galaxy is difficult. It is then useful to define the age of a galaxy as the time when it has built-up a significant fraction of its observed stellar mass. We therefore use two other age definitions that represent the time since the galaxy has built up 25\% and 50\% of its current stellar mass at the time of observation. We refer to these ages as the quarter mass age noted $\mathcal{A}_{M/4}$ and the half mass age, $\mathcal{A}_{M/2}$, respectively. Mathematically they are defined as:
\begin{eqnarray}
&  & \mathcal{A}_{M/4}= \mathcal{A}_{onset}-t_{M/4} \\
&\mathrm{with} &\;1/4\times\mathcal{M}_{\star,obs}=\int^{t_{\mathcal{M}/4}}_{0}SFH(t)dt, \nonumber \\
 &  & \mathcal{A}_{M/2}= \mathcal{A}_{onset}-t_{M/2}\\
&\mathrm{with} &\;1/2\times\mathcal{M}_{\star,obs}=\int^{t_{\mathcal{M}/2}}_{0}SFH(t)dt, \nonumber 
\end{eqnarray}
where $t_{M/4}$ and $t_{M/2}$ are the times where the quarter and the half of the stellar mass are built up and are represented in blue and red in Fig.\ref{Age-def-schema}, respectively.

Another common definition of age is the mass weighted age (MW for mass weighted age), used for example in cosmological simulations. This age is related to the relative importance of each population of stars that composes the galaxy. It is defined as the sum of the age of each of the population of stars in the galaxy weighted by their contribution to the stellar mass of the galaxy:
\begin{equation}
\mathcal{A}_{MW}=\sum_{n=1}^{N}\frac{\mathcal{A}_{onset,i}\mathcal{M}_{i}}{\mathcal{M}_{\,obs}},
\end{equation}
where $\mathcal{A}_{onset,i}$, $\mathcal{M}_{i}$ and $\mathcal{M}_{\star,obs}$ are the age of the \textit{i-th} population, its mass and the observed stellar mass of the galaxy. The parameter in this definition is the number of bins, or equivalently, the number of stellar populations present in the galaxy. We tested 3 different time binnings: 10, 100 and 1000. We find that above a binning of 100 mass waighted age results remain unchanged and we adopt this binning for this study. 

\begin{figure}[h!]
\centering
\includegraphics[width=8cm,height=8cm]{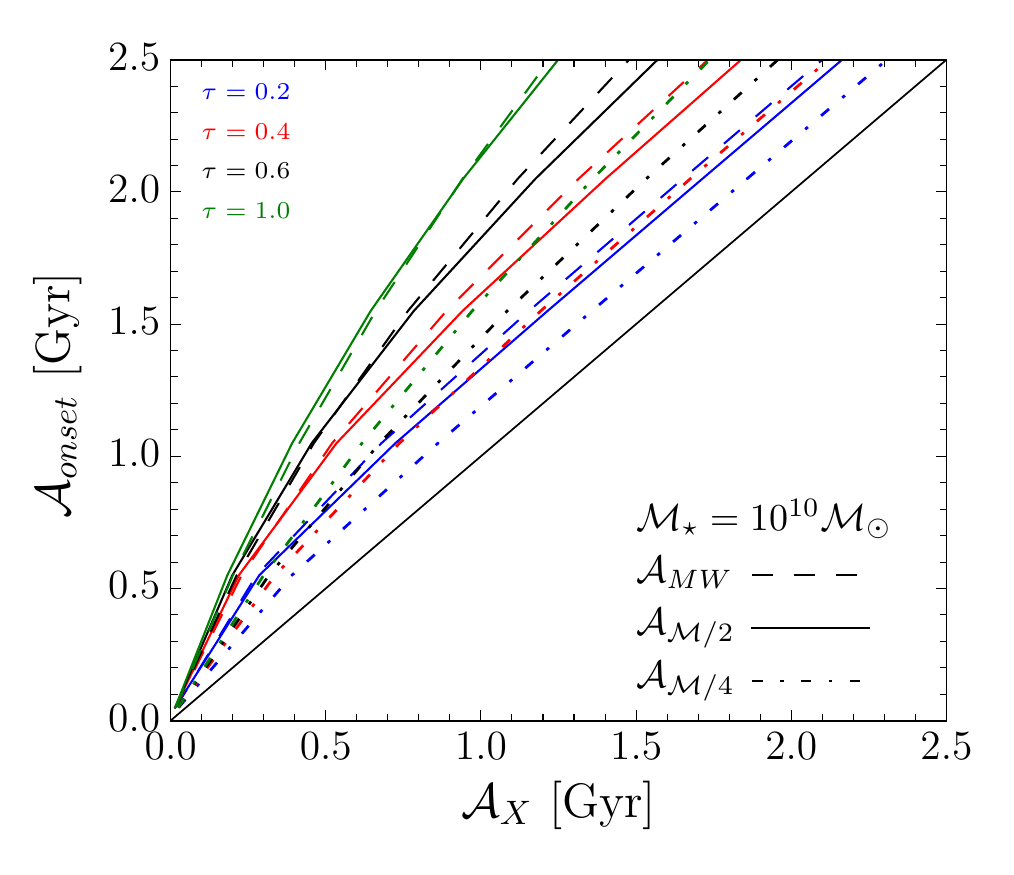}
\caption{Comparison of different age definitions to the age defined as the onset of star formation $\mathcal{A}_{onset}$. Different colours indicate different time-scales for the SFH: $\tau_{SFH}=0.4$ Gyr in red, $\tau_{SFH}=0.6$ Gyr in black, and $\tau_{SFH}=1.0$ Gyr in green.  The different line-styles correspond to three age definitions: solid line for half-mass ages ($\mathcal{A}_{\mathcal{M}/2}$), dashed line for mass weighted ages ($\mathcal{A}_{MW}$), and dot-dashed line for quarter-mass ages ($\mathcal{A}_{\mathcal{M}/4}$).}
\label{Age_def_iso_tau}
\end{figure}

Each age definition depends on the star formation history that is used to compute M$_{\star}$ and its time-scale parameter $\tau_{SFH}$ as shown in Figure \ref{Age_def_iso_tau}. The larger the time-scale, the larger the differences between different age definitions: as $\tau_{SFH}$ increases, the width of the star formation burst is getting broader and the time to accumulate one \textit{half} or one \textit{quarter} of the stellar mass is getting much larger. The quarter-mass age is evidently the closest to $\mathcal{A}_{onset}$ while the half -mass age shows a larger difference with $\mathcal{A}_{onset}$ since $t_{\mathcal{M}/2}>t_{\mathcal{M}/4}$. Figure \ref{Age_def_iso_tau} also shows that the half-mass age is very close to the mass-weighted age. Therefore, we do not use the half-mass age in the following but study only the mass weighted age, the quarter mass age and the age corresponding to the onset of the SFH, $\mathcal{A}_{onset}$.

The robustness against the spectro-photometric fitting with GOSSIP+ of each of these definitions ($\mathcal{A}_{onset} $, $\mathcal{A}_{\mathcal{M}/4} $ and $\mathcal{A}_{MW} $) is presented in the next Section.

\section{Measuring ages and degeneracies at $z>2$: Simulations}
\label{simul}

For the relatively young stellar populations expected in galaxies at high redshifts it is important to identify possible degeneracies between age measurements and other parameters. 

At low redshifts, the main degeneracies linked to the age parameter are the age-metallicity degeneracy and the age-dust degeneracy (see Section \ref{degsec}). At $z>2$ degeneracies are expected to be less severe than at lower redshifts as the age of a galaxy is strongly restricted to be smaller than the age of the Universe at the observed redshift. Assuming the parameters of the cosmological world model, the ages of galaxies are limited to a maximum of $\sim$0.8 Gyr at $z=6$ and $\sim $ 3.3 Gyr at $z=2.0$, the redshift boundaries of our sample. 

In the following we perform extensive simulations of galaxies with similar properties to the VUDS sample, and using spectroscopy and photometry jointly when performing model fitting. We analyse possible degeneracies that could affect age estimations and conclude on the accuracy of age measurements at these redshifts.

\subsection{Simulations: 180\,000 galaxies representative of VUDS properties}

To test the accuracy and robustness of age measurements of galaxies at z$>$2 we produce a large set of mock galaxies. Then, we run the SED fitting with GOSSIP+ to measure their ages, and compare them with input ages in the simulation. 

We simulate both broadband photometry and the associated spectra covering the same wavelength range as VUDS observations ($3650<\lambda<9350$\AA). We create a sample of mock galaxies with the same properties as our VUDS sample ($2<z<6.5$, $22<$i$_{AB}<25$ and signal-to-noise ratio $2<SNR<25$), i.e., with the same redshift, magnitude $i_{AB}$ and SNR combinations. For each of these VUDS combinations, 50 mock galaxies are created, each with a different galaxy synthetic model. The final simulated sample contains ~180\,000 objects. 


The synthetic model is randomly chosen in a BC03 library composed of 10\,000 templates (with $400\AA<\lambda<60000\AA$ restframe). In this library, the dust extinction is computed from the Calzetti law and E(B-V) values range from 0.0 to 0.5 (5 equal steps). The SFH is an exponentially delayed model (containing both increasing and declining SFH sections) with a timescale $\tau_{SFR}$  (which represents the time between the onset of the star formation and the peak of the SFH) that ranges from 0.1 to 5.0 Gyr (in 10 steps). We use the Chabrier IMF \cite{Chabrier2003}. Ages are allowed to be in the interval [0.05;4.0] Gyr (0.05 Gyr steps between 0.05 Gyr and 2.5 Gyr and 0.25 above), and the selected model has to be younger than the age of the Universe at the redshift of the galaxy. The metallicity is in the range [Z=0.004; Z=0.05] (4 values, Z$_{\odot}$=0.02). Finally, an IGM template is chosen randomly among 7 possibilities at the considered redshift (see Section \ref{GOSSIP+} and \cite{Thomas14} for details). 

After the selection of a triplet ($z$,$i_{AB}$,SNR) we randomly select a synthetic model, redshifted at a redshift $z$. Then, the IGM transmission is applied. The model is normalised to $i_{AB}$ and re-sampled to the same spectral resolution as the observations. The wavelength of the final mock spectra ranges from 3600\AA~to 9500\AA~ with a $\Delta\lambda=5.35$\AA~ to mimic the VUDS spectra. Poisson noise is then added to the data as the square root of the simulated flux.
Finally, the magnitudes are computed by convolving the model with the transmission of several photometric filters: \textit{ugriz} from the Megacam camera, JHK from the WIRCAM camera and two Spitzer IRAC band at 3.6 and 4.5$\mu m$.

\subsection{Fitting process}

Once the simulations are created we run GOSSIP+ three times: on spectra alone, on the photometry alone, and on both combined. For this process we use two different fitting libraries one with the BC03 population synthesis model and the other with M05. The parameter space used for the two libraries is presented in table \ref{Paramspace_simufit}.

\begin{table}[h!]
\caption{Parameter space used during the fit of the mock galaxies with both BC03 and M05 models.}
\label{Paramspace_simufit}    
\centering                       
\begin{tabular}{c c c }       
\hline\hline                
Parameter & BC03 & M05 \\   
\hline                       
IMF & Chabrier & Chabrier \\ 
Metallicity [Z] &  0.004,0.008,0.02,0.05 & 0.001,0.01, 0.02, 0.04 \\ 
$\tau_{SFH}$ & \multicolumn{2}{c}{0.1 to 5.0 Gyr (10 steps)}  \\
 E(B$-$V) & \multicolumn{2}{c}{0.0, 0.1, 0.2, 0.3, 0.4, 0.5}\\ 
Ages (Gyr) & \multicolumn{2}{c}{0.05 to 4.0 (30 steps)}  \\ 
IGM & \multicolumn{2}{c}{7 templates per redshift} \\ 
\hline                                  
\end{tabular}
\end{table}

This provides a large statistical basis to examine the robustness of the age computation, and various possible associated degeneracies. We present the results in the next two subsections, with first a visual inspection of which degeneracies are present, followed by a quantitative analysis of age estimates.

\subsection{Stellar mass and Star formation rate}

Before focusing on age measurements in the next Section we discuss here the reliability of GOSSIP+ in measuring M$_{\star}$ and SFR from the combined fitting of spectra and broad-band photometry. We use the simulations as defined above and we compare the M$_{\star}$ and SFR from the three types of fit, on the set of simulated broad-band photometric magnitudes alone, on simulated spectra only and on the combined photometry and spectroscopy datasets.

Figures \ref{Mass_maps} and \ref{SFR_maps} compares M$_{\star}$ and SFR obtained from each type of fit with the input values of the simulation. In both Figures, the top panel shows the density map of combined data with its $1\sigma$ contour, while the bottom panel, shows the evolution of $\Delta  \mathrm{Param}= \mathrm{Param}_{in}-\mathrm{Param}_{out}$ as a function of the parameter $Param$ for the three types of fits. Table \ref{SIMU_Mass_SFR_table} gives the measurement of $\Delta  \mathrm{Param}$ for each parameter combination and fit type for the full simulation, with the evolution of this quantity as a function of redshift.

\subsubsection{Stellar mass, M$_{\star}$}

\begin{figure}[h!]
  \centering
  \includegraphics[width=\hsize]{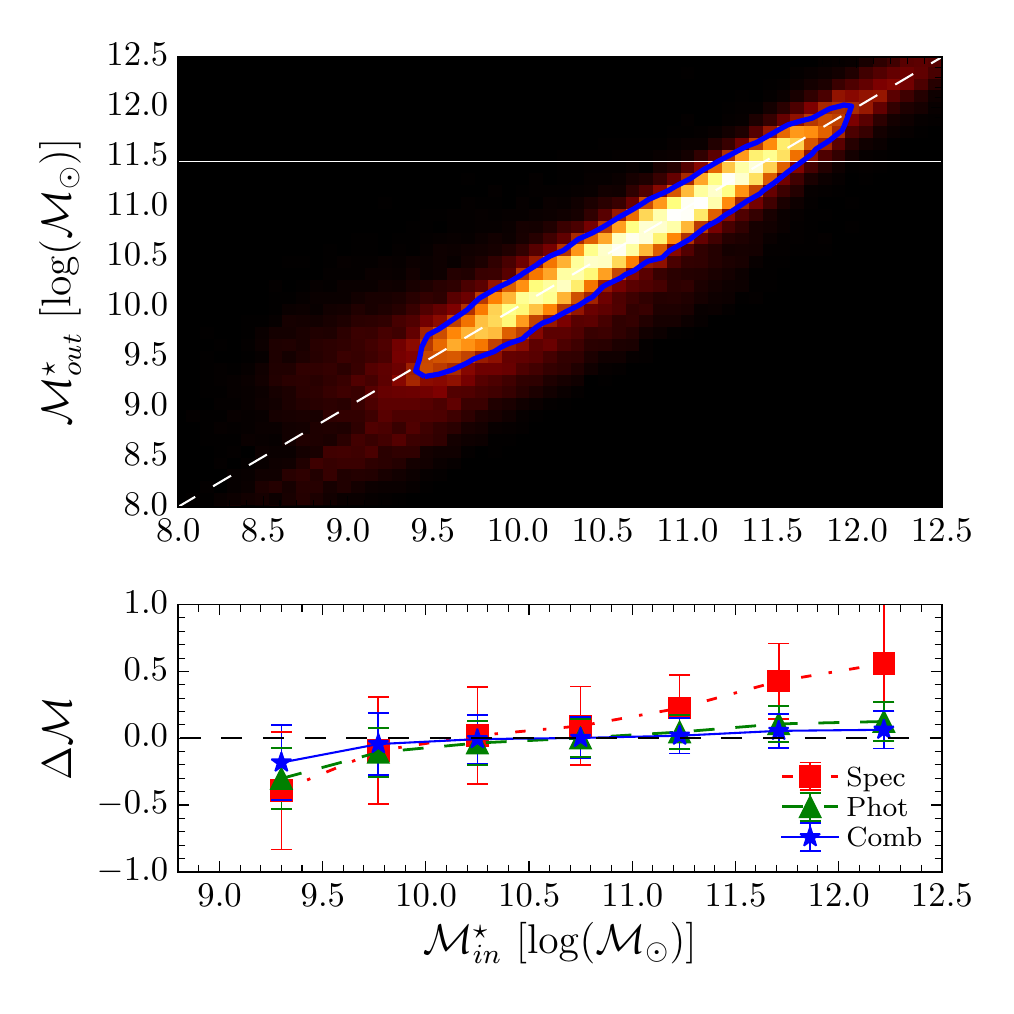}
  \caption{Comparison of the stellar mass measurements M$^{\star}_{out}$ obtained using GOSSIP+ on the 180\,000 simulated M$^{\star}_{in}$ sample with the input values. \textit{Top panel}: The density plot represents the measurement of M$_{\star}$  from the combined fit versus the input M$_{\star}$. The highest density is represented by the brightest colour, while the lowest density is represented in black. The blue line represents the 1$\sigma$ contour of the distribution. The combined fit is able to find the correct M$_{\star}$ at any mass in the simulated range. \textit{Bottom panel}: Evolution of the quantity $\Delta \mathcal{M}= \mathcal{M}^{\star}_{in}-\mathcal{M}^{\star}_{out}$ over the stellar mass range. The points in red, green and blue represent the fit on the spectroscopy only, photometry only and on the combined data, respectively. The combined fit leads to the most accurate M$_{\star}$ computation with $\Delta M_{\star}<0.1$ dex.}%
  \label{Mass_maps}%
\end{figure}

The comparison of M$_{\star}$ computed with GOSSIP+ to the simulated mass is shown in Figure \ref{Mass_maps} and Table \ref{SIMU_Mass_SFR_table} for the three cases considered here. 

The spectroscopic data covering the UV rest-frame as in VUDS does recover M$_{\star}$ on average, but with a relatively poor accuracy. The difference between the input and measured values at $\log_{10}$M$_{\star}>10.2$ is $\Delta \mathcal{M}_{\star} \sim 0.3$dex on average while at $\log_{10}$M$_{\star}<10$ it is $\sim-0.20$ dex. The median absolute deviation is of $\sim 0.4$ dex.  
With UV rest-frame only the spectroscopic data obviously do not constrain the NIR which is most sensitive to the stellar mass. Therefore, the resulting errors on M$_{\star}$ are large, as shown by the evolution of $\Delta \mathcal{M}$ particularly for high mass galaxies. This is due to the fact that high mass galaxies generally host older and then redder stellar populations, difficult to constrain from UV rest-frame spectroscopy only. As expected, the UV rest-frame spectroscopy alone is not the right observable to constrain the stellar mass. 

The fit on the photometric data alone provides M$_{\star}$ measurements with good accuracy. This is widely used in the literature e.g. from photometric redshift codes (\citealt{Ilbert13} and reference therein). We note that the accuracy in measuring M$_{\star}$ slightly decreases with redshift, going from -0.09 dex at z$\sim$2.4 to -0.21 dex at $z>4.25$. This can be explained by the fact that at high redshifts, the NIR photometry corresponds to the UV rest-frame part of the data, then imposing fewer constraints on the older stellar populations tracing the stellar mass. GOSSIP+ is able to compute M$_{\star}$ from a photometric SED as reliably as most standard codes in the literature.

The fit on the combined spectroscopic and photometric data shows an excellent accuracy in M$_{\star}$ measurements, at all redshifts. At most redshifts explored here the accuracy in M$_{\star}$ is typically $\sim$0.02 dex, almost five times better than the photometry alone at $z\sim2$ and always twice better at any higher redshift. This can be explained by the fact that these data combine the IR constraints from the photometry with the UV rest-frame from spectroscopy leading to a finer separation of model templates in the parameter space explored. The evolution of $\Delta \mathcal{M}$ from the fit on combined data shows that for low mass galaxies, the addition of the photometry allows to limit an overestimation of the mass, while for high mass galaxies we observe the opposite behaviour. 

This analysis on the stellar mass shows that the combined fit of broad-band photometric data covering from the u-band to 4.5$\mu$m with UV rest-frame spectroscopic data leads to a significant improvement in the estimation of this important physical parameter.

\subsubsection{SFR}

\begin{figure}[h!]
  \centering
  \includegraphics[width=\hsize]{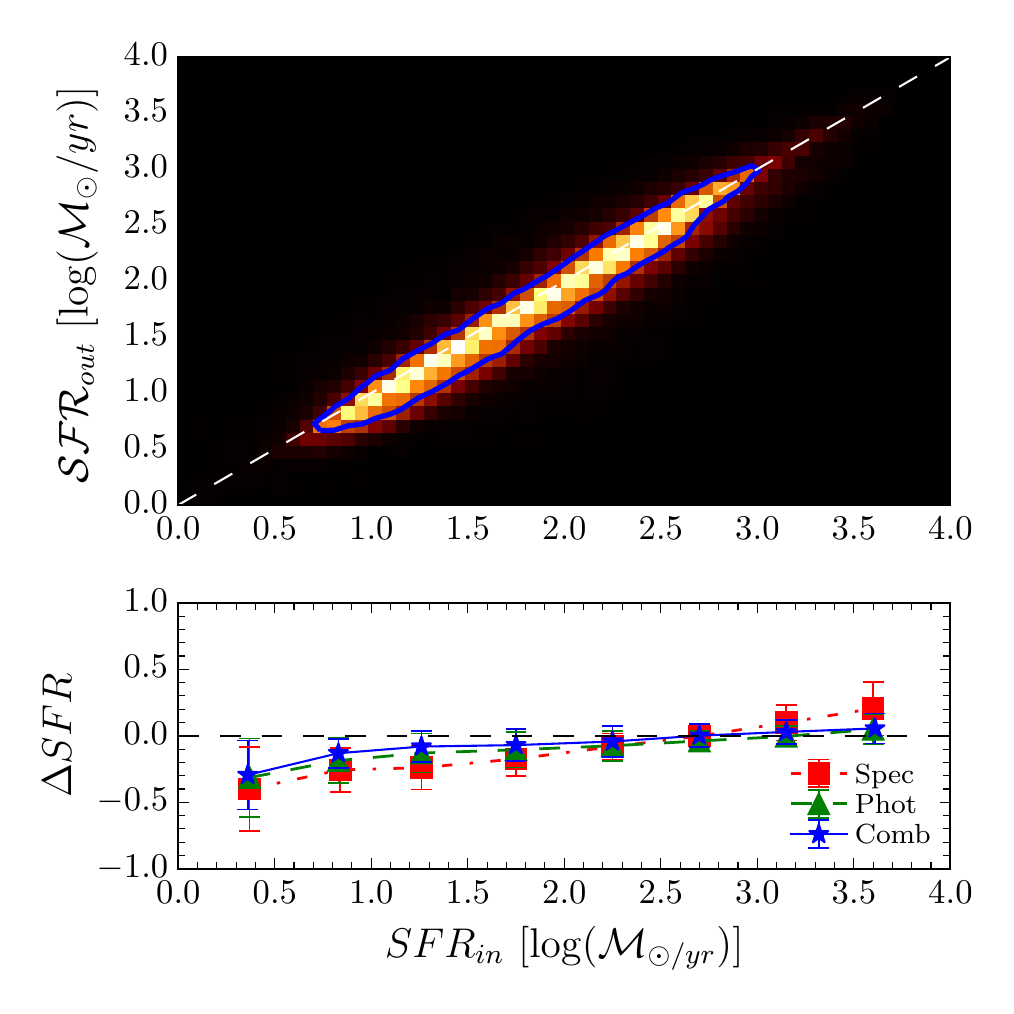}
  \caption{Comparison of SFR recovered using SED fitting with GOSSIP+ to the simulated dataset of 180\,000 galaxies. colours and line types are as for Figure \ref{Mass_maps}}%
  \label{SFR_maps}%
\end{figure}

Figure \ref{SFR_maps} and Table \ref{SIMU_Mass_SFR_table} show the estimation of the SFR against the input values of our 180\,000 simulated galaxy sample. For this parameter the three types of fit give closer results than for M$_{\star}$. 

The fit on the spectroscopic data alone is the least accurate one with an accuracy of 0.2 dex on average. The accuracy improves at all redshifts when using the photometric data alone. The estimation of the SFR with the fit on the spectroscopy and photometry combined is particularly efficient up to $z\sim4.25$,  with an average $\Delta \mathcal{SFR} \sim -0.15$dex and a median absolute dispersion of $-0.13$. The combination of UV rest-frame spectroscopic data, sensitive to recent star formation, with optical and NIR broad-band photometric data which is sensitive to star-formation cumulated over a long time baseline, is particularly efficient in recovering the SFR. Going to higher redshifts, the accuracy is becoming worse at $\sim$-0.18 dex since half of the observed spectrum is composed by the region below the Lyman limit which does not bring constraints on the SFR, but the combined fit recovers the SFR more accurately than the fit on spectroscopy or photometry taken separately. 

\begin{table*}[h]
\caption{Estimation of the stellar mass and star formation rate from the three types of fit. For each combination of parameter and fit type, we give the mean of the quantity $\Delta  \mathrm{Param}= \mathrm{Param}_{in}-\mathrm{Param}_{out}$ and the associated median absolute deviation.}
\label{SIMU_Mass_SFR_table}    
\centering                       
\begin{tabular}{c c c c c c c }       
\hline\hline                
Parameter/Fit & Full simulation & $ 2.0 \leq z < 2.75 $ & $2.75  \leq z <  3.5 $ & $ 3.5 \leq z < 4.25 $ & $ z \geq 4.25 $  \\   
\hline\hline                        
Mass/SPEC &  0.18 $\pm$ 0.39   &   0.21 $\pm$ 0.39 &  0.20 $\pm$ 0.39 &  0.19 $\pm$ 0.40 &  0.16 $\pm$ 0.49 \\ 
Mass/MAGS & -0.09 $\pm$ 0.18   &  -0.05 $\pm$ 0.16 & -0.11 $\pm$ 0.19 & -0.11 $\pm$ 0.21 & -0.21 $\pm$ 0.28\\ 
Mass/COMB & \textbf{-0.02} $\pm$ \textbf{0.19}   &   \textbf{0.01} $\pm$ \textbf{0.16} & \textbf{-0.03} $\pm$ \textbf{0.20} & \textbf{-0.05} $\pm $\textbf{0.21} & \textbf{-0.12} $\pm$ \textbf{0.27} \\
\hline
\hline
SFR/SPEC &  -0.20 $\pm$ 0.13   &  -0.20 $\pm$ 0.12 & -0.20 $\pm$ 0.14 & -0.18 $\pm$ 0.16 & -0.14 $\pm$ 0.26\\ 
SFR/MAGS &  -0.19 $\pm$ 0.14   &  -0.16 $\pm$ 0.13 & -0.17 $\pm$ 0.12 & -0.13 $\pm$ 0.13 & -0.20 $\pm$ 0.19\\ 
SFR/COMB &  \textbf{-0.13} $\pm$ \textbf{0.12}   &  \textbf{-0.14} $\pm$ \textbf{0.12} & \textbf{-0.13} $\pm$ \textbf{0.13} & \textbf{-0.10} $\pm$ \textbf{0.11} & \textbf{-0.18} $\pm$ \textbf{0.16}\\ 
\hline                                  
\end{tabular}
\end{table*}

\subsection{Estimation of galaxy ages, quantitative analysis}

In this Section we analyse our simulations to assess the robustness of age measurement for the three types of fit, following a similar methodology as for M$_{\star}$ and SFR in the previous Sections. We analyse results for the 3 definitions of age presented in Section \ref{sec_age_def}, $\mathcal{A}_{onset}$, $\mathcal{A}_{\mathcal{M}/4}$ and $\mathcal{A}_{MW}$. We therefore create in both input (output) the mass-weighted age and half-mass age from the input (output) SFH and $\mathcal{A}_{onset}$ as described in section \ref{sec_age_def}. 

\begin{figure}[h!]
  \centering
  \includegraphics[width=\hsize]{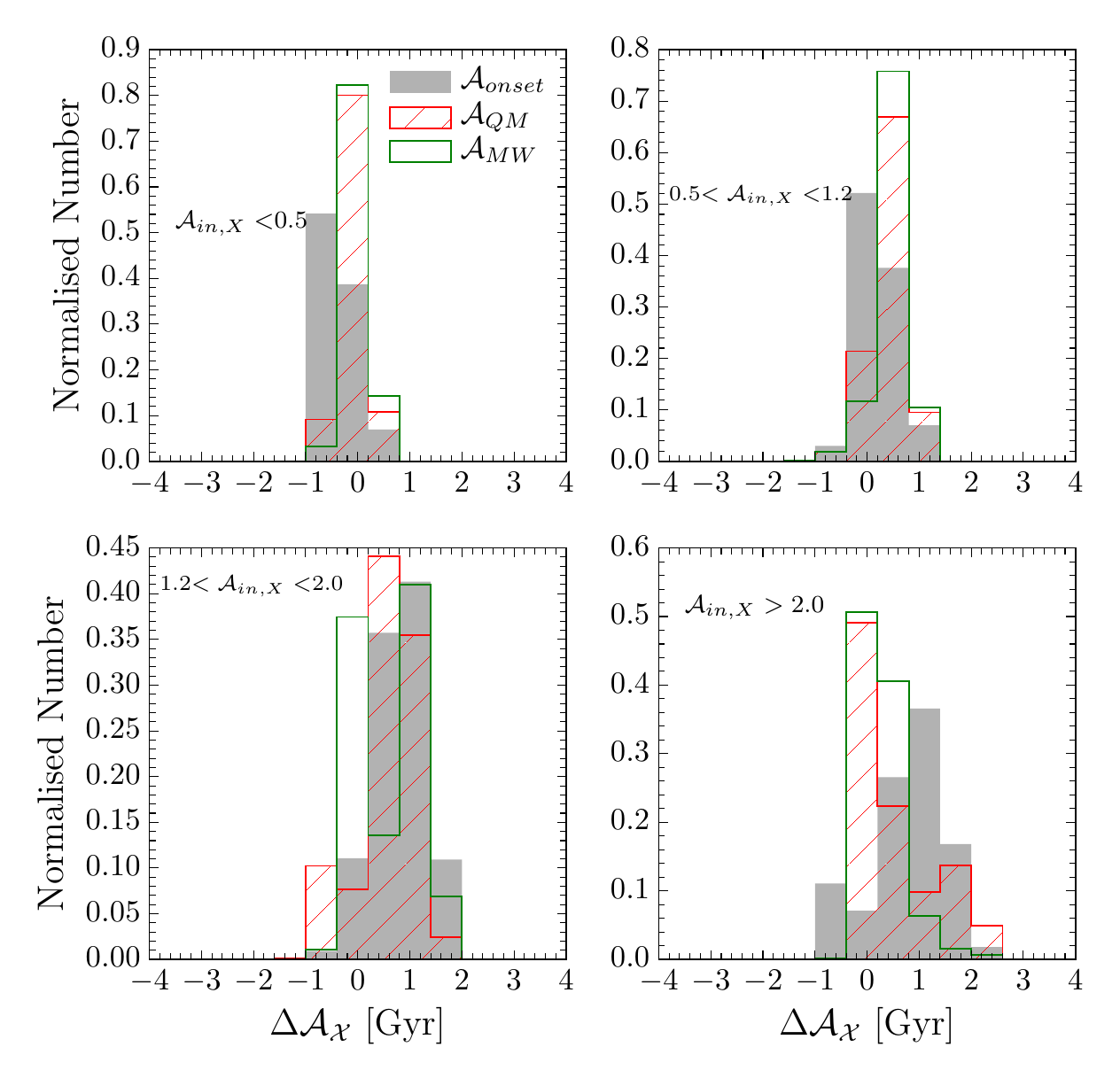}
  \caption{Comparison of the $\Delta \mathcal{A}$ for the three definitions of ages from the fit on the spectroscopy only. In grey, we show the $\mathcal{A}_{onset}$ definition, in green $\mathcal{A}_{MW}$ and in red $\mathcal{A}_{\mathcal{M}/4}$. We show three input age bins. From top left to bottom right, $\mathcal{A}_{in}<0.5$, $0.5<\mathcal{A}_{in}<1.2$, $1.2<\mathcal{A}_{in}<2.0$ and $\mathcal{A}_{in}>2.0$. All the ages are given in Gyr.}%
  \label{Age_spectro_simu}%
\end{figure}

\begin{figure}[h!]
  \centering
  \includegraphics[width=\hsize]{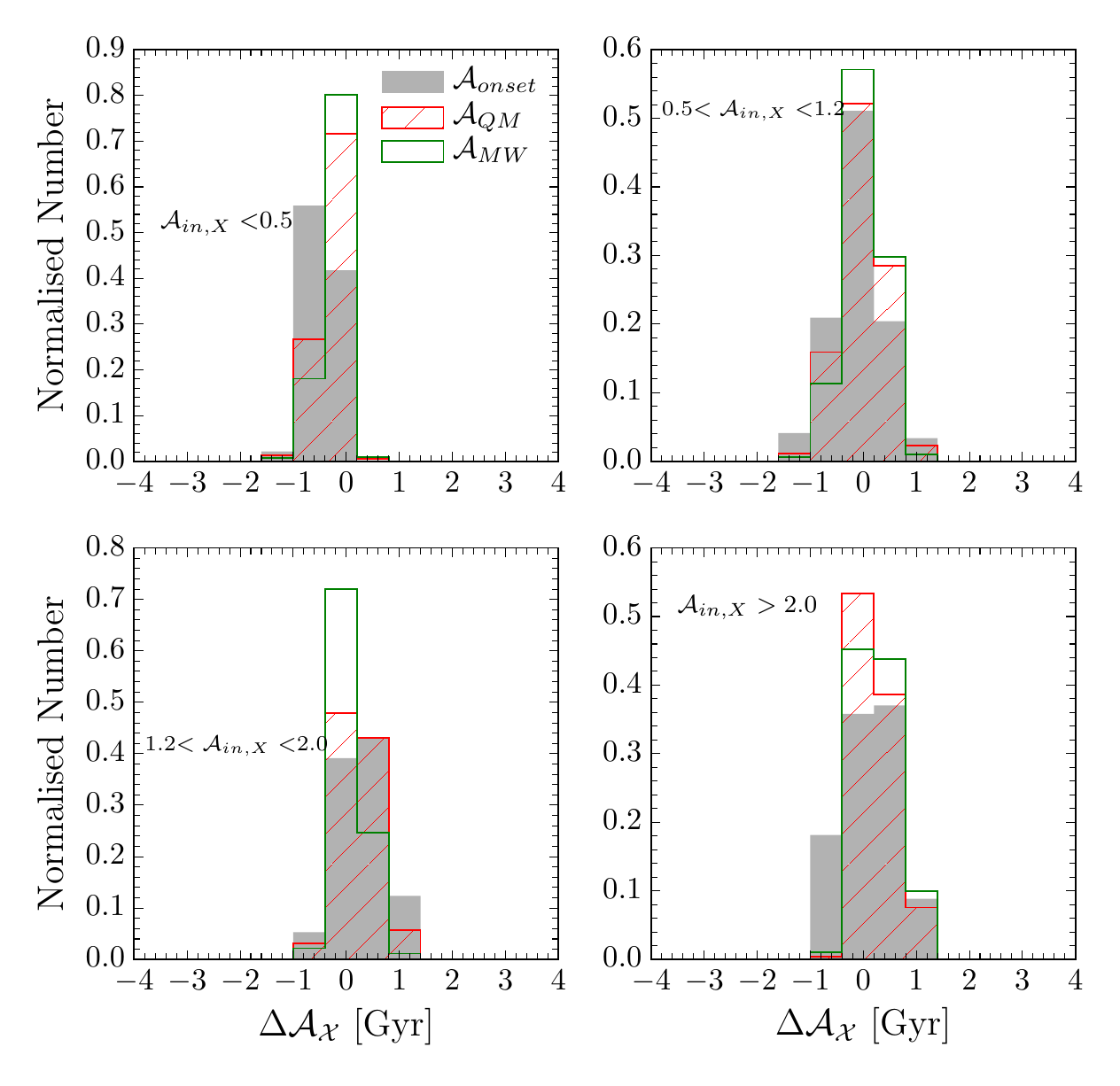}
  \caption{Same as \ref{Age_spectro_simu} but for the fit on the photometry only.}%
  \label{Age_photo_simu}%
\end{figure}

\begin{figure}[h!]
  \centering
  \includegraphics[width=\hsize]{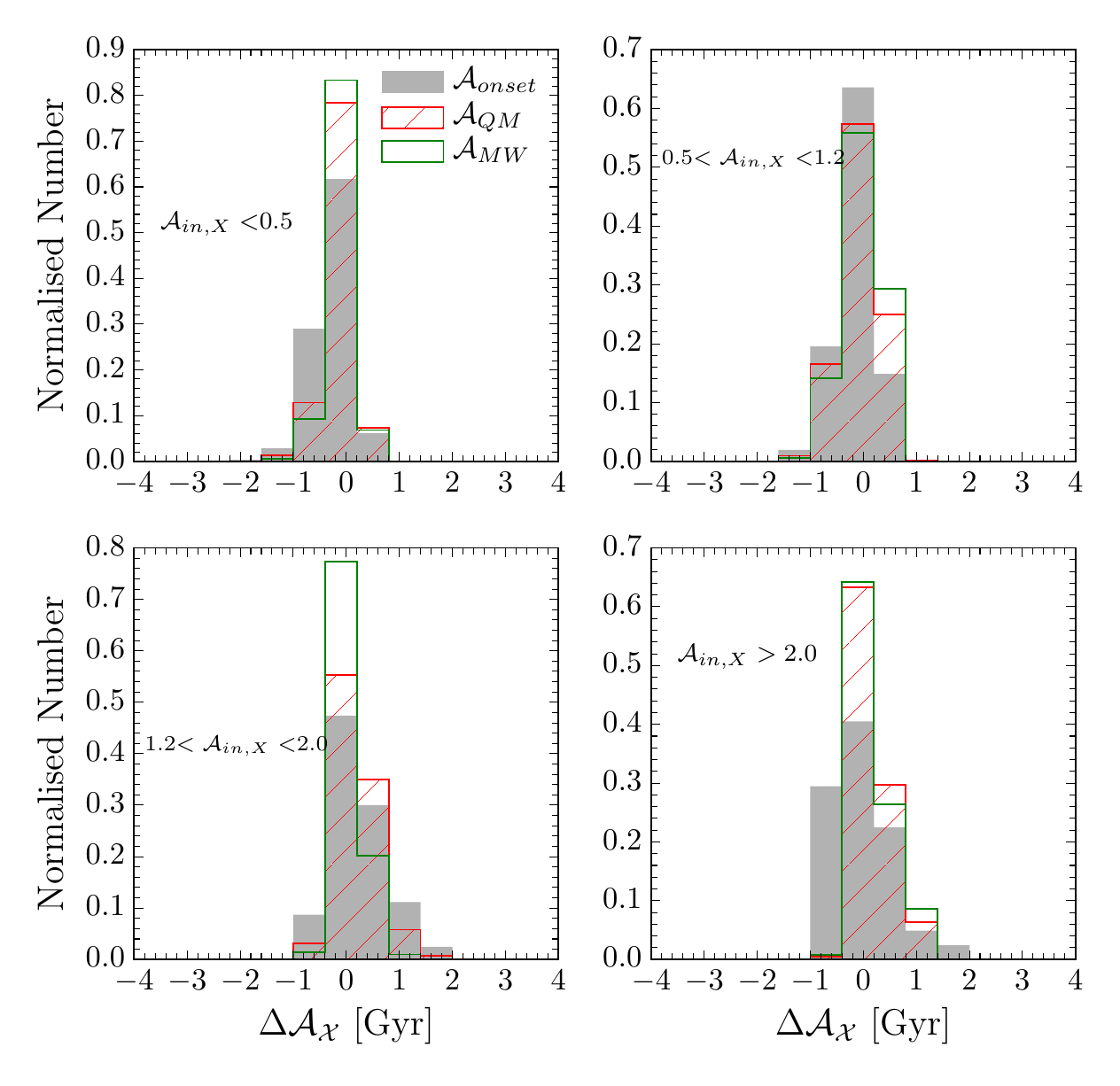}
  \caption{Same as \ref{Age_spectro_simu} but for the combined fit of spectroscopy and photometry.}%
  \label{Age_comb_simu}%
\end{figure}

Results are presented from Figures \ref{Age_spectro_simu} to \ref{Age_comb_simu} and in Table \ref{SIMU_Age_table}. In each Figure, we present the $\Delta\mathcal{A}$ in four $\mathcal{A}_{in}$ bins, and for the three age definitions.

Not surprisingly, the fit on the spectroscopy alone provides the worst estimate of galaxy ages. Using $\mathcal{A}_{onset}$, the difference between the input and the measured ages on the whole simulation is around 0.3 Gyr and the median absolute deviation is of 0.50 Gyr. The age is generally underestimated. The accuracy is improving with redshift as expected since the upper age limit is given by the age of the Universe and therefore decreases with redshift. This is confirmed from the evolution of $\Delta Age$ as presented in Figure \ref{Age_spectro_simu} as a function of the input age. When the simulated age is higher than 2.0 Gyr (only at redshift $z\lesssim3$) $\Delta Age$ is 1.0 Gyr, but $\Delta Age \sim 0.25$ when the input age is smaller than 1.2 Gyr (corresponding to a redshift $z\gtrsim4.5$). 

For $\mathcal{A}_{\mathcal{M}/4}$  we note that $\Delta Age$ is slightly improved with respect to the $\mathcal{A}_{onset}$ definition. This is particularly the case for the highest input age where the accuracy reaches 0.5 Gyr against 0.8 for $\mathcal{A}_{onset}$. This is similar for the median absolute deviation. This behavior is expected since ages are artificially reduced in this definition, leading to lower value of $\Delta Age$ and median absolute deviations. Finally, $\mathcal{A}_{MW}$ gives a better estimation of the age with $\Delta Age \sim 0.16$ Gyr for the full simulation. As shown in Section \ref{sec_age_def}, this definition gives the smallest age, leading to a smaller $\Delta Age$.

Using the photometry alone (Fig. \ref{Age_photo_simu}) improves the estimation of the age compared to the use of the spectroscopic data only. The fit is better constrained over a large wavelength range by the full set of photometric point the NIR and IR regions provide constraints on the age of the oldest stars. The difference in age for $\mathcal{A}_{onset}$ is typically $\Delta Age \sim 0.15$ Gyr, with a median absolute deviation of 0.33 Gyr. The age parameter with the lowest $\Delta Age$ is the mass-weighted-age. As described in the previous paragraph, the artificial reduction of the age with this definition leads to an artificial reduction of the results of the simulation. For this parameter $\Delta Age = -0.08$ Gyr.

Combining photometry and spectroscopy clearly provides the best age estimates, for any of the three age definitions considered (Fig. \ref{Age_comb_simu}). A strong constraint on young massive stars is imposed by the UV rest-frame spectroscopic data and the optical and NIR photometric data constrains the population of oldest stars. For $\mathcal{A}_{onset}$ the age difference between the simulated input and the measurements is better than $\Delta Age \sim 0.1$ Gyr for the whole simulation  with a median absolute deviation of 0.32 Gyr. The age difference between input and output is the smallest for the mass weighted age $\mathcal{A}_{MW}$, in agreement with the findings of \citet{Pforr12} at all redshifts: the average $\Delta Age$ is 0.03 Gyr with a median absolute deviation of 0.15 Gyr. 
This level of accuracy in measuring galaxy ages at these high redshifts is unprecedented. 

To summarize, this analysis demonstrates that the combined use on high redshift galaxies (z>2) of high quality UV rest-frame spectra and deep multi-wavelength photometry including rest-frame optical and NIR data allows to significantly improve galaxy age measurements. This combination of different information allows to increase the accuracy by a factor 2 for the $\mathcal{A}_{onset}$ definition with respect to the classical SED-fitting and by factor more than three with respect to the fit on the spectroscopic data. For the parameter that gives the smallest age, $\mathcal{A}_{MW}$, the combined fit shows an accuracy almost 3 times better than the photometric fit and more than 5 times better the the spectroscopic fit. Age measurements using this method show an uncertainty of $\sim9-10$\% ($1\sigma$) comparable to the uncertainty in M$_{\star}$ measurements.

With this level of accuracy it becomes possible to investigate age-related properties to get a complementary point of view on galaxy evolution compared to using a more restricted analysis of M$_{\star}$ or SFR alone.  

\begin{table*}[h]
\caption{Estimation of galaxy ages for different age definitions from the three types of fit. For each combination of parameter and fit type, we give the mean of the quantity $\Delta  \mathrm{Param}= \mathrm{Param}_{in}-\mathrm{Param}_{out}$ and the associated median absolute deviation.}
\label{SIMU_Age_table}    
\centering                       
\begin{tabular}{cccccc}
\hline 
\hline 
Age definition / Fit & Full sample & $2<z\leq 2.75 $ & $2.75 < z \leq 3.5$ &  $3.5<z\leq 4.25 $& $z\geq4.25$ \\ 
\hline
\hline  
$\mathcal{A}_{onset}$/SPEC & 0.28 $\pm$ 0.50 & 0.33$\pm$0.61 & 0.28$\pm$0.50 & 0.25$\pm$0.40 & 0.23$\pm$0.34  \\ 
$\mathcal{A}_{onset}$/MAGS & 0.14 $\pm$ 0.33 & 0.18 $\pm$ 0.37 & 0.14 $\pm $ 0.32& 0.12 $\pm$ 0.29 & 0.10 $\pm$ 0.28  \\ 
$\mathcal{A}_{onset}$/COMB & \textbf{0.08} $\pm$ \textbf{0.32} & \textbf{0.14} $\pm$ \textbf{0.34} & \textbf{0.07} $\pm$ \textbf{0.31} & \textbf{0.05} $\pm$ \textbf{0.27}& \textbf{0.04} $\pm$ \textbf{0.29}  \\ 

\hline
\hline
$\mathcal{A}_{\mathcal{M}/4}$/SPEC& 0.25 $\pm$ 0.32 & 0.29 $\pm$ 0.39 & 0.25$\pm$0.33 & 0.18$\pm$0.25 & 0.15$\pm$0.23  \\ 
$\mathcal{A}_{\mathcal{M}/4}$/MAGS& -0.12 $\pm$ 0.22& -0.16 $\pm$ 0.24& -0.12 $\pm$ 0.21 & -0.13 $\pm$ 0.17  &  -0.13$\pm$0.18 \\ 
$\mathcal{A}_{\mathcal{M}/4}$/COMB& \textbf{-0.04}$\pm$\textbf{0.20} &\textbf{-0.09}$\pm$\textbf{0.19}  & \textbf{-0.005}$\pm$\textbf{0.18}  & \textbf{0.00}$\pm$\textbf{0.19}  &\textbf{0.04}$\pm$\textbf{0.18}   \\ 

\hline
\hline
$\mathcal{A}_{MW}$/SPEC& 0.16 $\pm$0.23 & 0.19$\pm$0.29 & 0.17$\pm$0.23  & 0.13$\pm$0.18 &0.11$\pm$0.16    \\ 
$\mathcal{A}_{MW}$/MAGS& -0.08 $\pm$ 0.17 & -0.10$\pm$0.20 & -0.07$\pm$0.17 & -0.07$\pm$0.13 & -0.07$\pm$ 0.14  \\ 
$\mathcal{A}_{MW}$/COMB& \textbf{0.03}$\pm$\textbf{0.15} & \textbf{-0.04}$\pm$\textbf{0.16} & \textbf{0.03}$\pm$\textbf{0.15}  & \textbf{0.02}$\pm$\textbf{0.13} & \textbf{0.04}$\pm$\textbf{0.13}   \\ 
\hline 
\end{tabular} 
\end{table*}

\subsection{Studying age-related degeneracies}
\label{degsec}
\subsubsection{Method}
Degeneracies are the major problem when computing galaxy ages. To study what is the influence of both the age limit imposed by the age of the universe and the addition of spectroscopy to photometry we developed a method based on probability distribution functions (PDFs). 
We created two dimensional PDFs, called \textit{probability density maps} (PDM), for any parameter in combination to age (we study here only $\mathcal{A}_{onset}$, but the conclusions derived for this age definition are qualitatively similar for the other age definitions). 

A PDM represents a probability density in 2D-space and  clearly highlights how the PDFs of the two parameters relate to each other and the possible degeneracies that are at play. PDMs are created as follows. For two parameters ($\delta_{1}$; $\delta_{2}$) which can take $N_{1}$ and $N_{2}$ values, the 2D-space is build with $N_{1}\times N_{2}$ points. During the fit, a $\chi^{2}$ value is assigned to each model template in the library and is used to derive a probability for the goodness of fit (see Equation \ref{proba_G}). Therefore, the probability $P$ of a given point in the PDM ($\delta_{1,i}$; $\delta_{2,j}$ for example) corresponds to the sum of the probability $p$ of the template that satisfies $\delta{1}=\delta_{1,j}$ and $\delta_{2}=\delta_{2;j}$: 

\begin{equation}
P(\delta_{1}; \delta_{2})=\sum_{k} p[template_{k}(\delta_{1}=\delta_{1,j}:\delta_{2}=\delta_{2;j})].
\end{equation}
Since the two-parameter space of the different PDMs is not continuous, the PDMs are linearly interpolated between parameter values (see next Section for numerous examples). To compare the quality of the fits we analyse three quantities:
\begin{itemize}
\item N$_{\mathrm{peak}}$: The number of peaks included in the $1\sigma$ contour. This indicates remaining degeneracies as each peak could host the right parameter values.
\item The position of the input parameters inside or outside the $1\sigma$ contour. Obviously if the input parameters are in the $1\sigma$ contour this indicates that the fit has properly converged towards the correct value.
\item The ratio of the area of the $1\sigma$ contour between two different fits (spectrum only, photometry only, or the two combined). Ratios smaller than one indicate an improvement of the fit quality because the parameter space allowed by the fit is reduced. Since maps are constructed on the same grid, the area of the  $1\sigma$ contour is calculated as the number of points (parameter pairs) that fall in this contour.
\end{itemize}
We study in the next two subsections the age--dust and age--metallicity degeneracies using the PDM formalism.
\subsubsection{Age-dust degeneracy}
One of the main difficulties in measuring ages is the age-dust degeneracy. Dust strongly attenuates flux at $\lambda \la 4500$\AA\ (rest-frame), re-emitting it in the far infrared. A strongly star forming galaxy with a large amount of dust will have a spectrum where the blue part is strongly attenuated and a bright red part, which may mimic old stellar populations and hence an old galaxy. From the PDMs in our simulations, we find that fitting spectroscopy or photometry alone is not able to reproduce the right pair of age and dust parameters. Both fit modes exhibit rather large 1$\sigma$ contours in the PDMs enclosing young age values with high dust content as well as older ages with lower E(B-V). 

\begin{figure*}[h!]
  \centering
  \includegraphics[width=15cm,height=6cm]{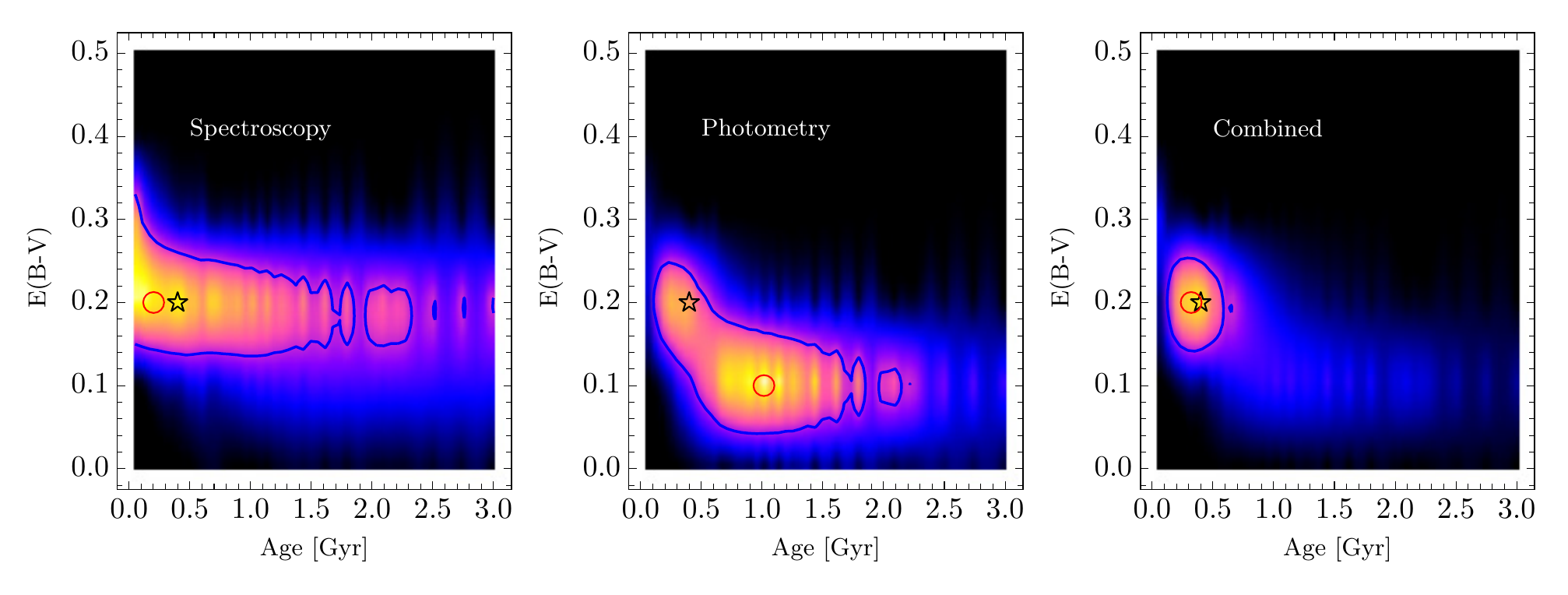}
  \includegraphics[width=15cm,height=6cm]{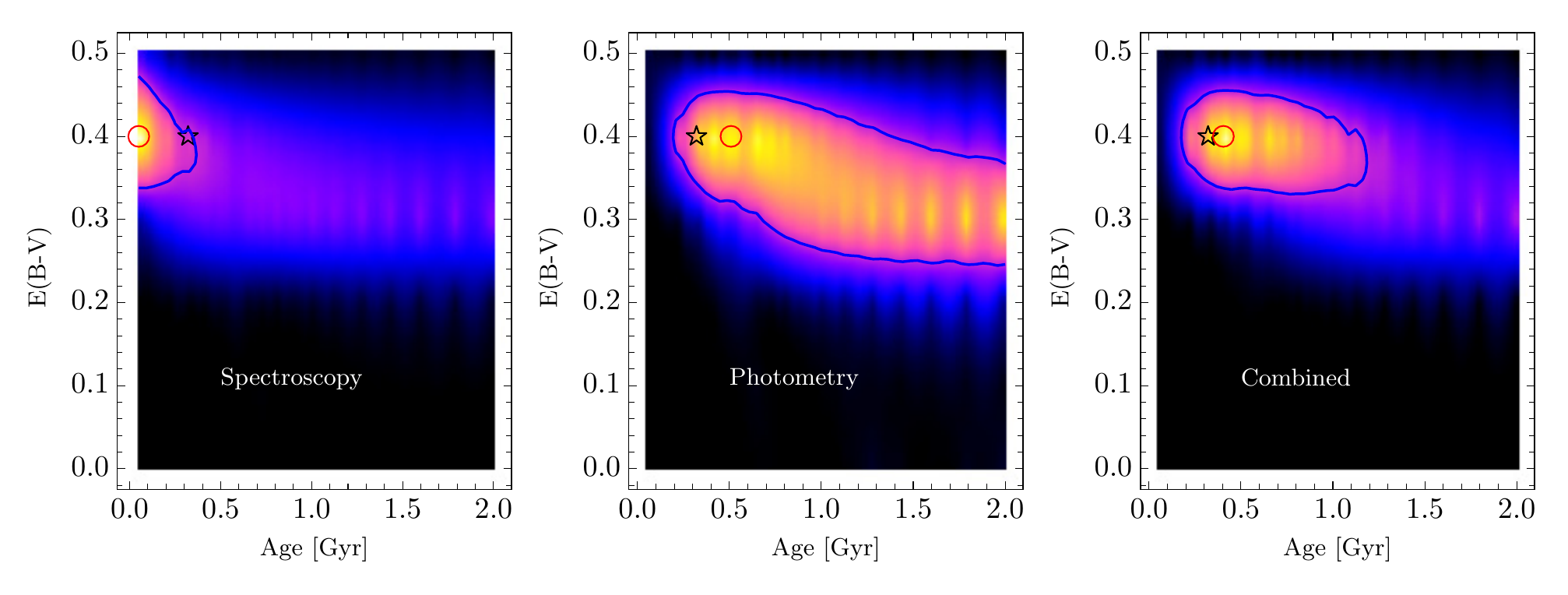}
   \includegraphics[width=15cm,height=6cm]{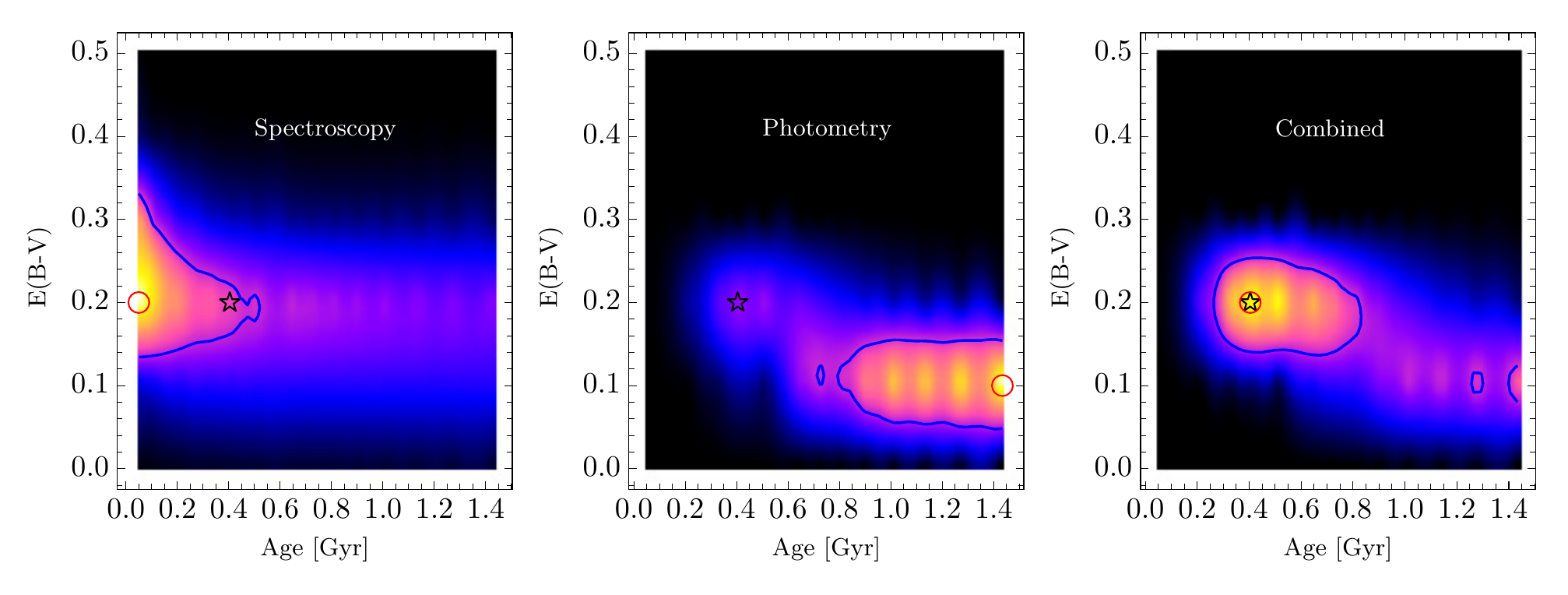}
  \caption{Nine examples of Probability density maps in our simulation. \textit{From left to right}: Fitting the photometry only, the spectroscopy only and the combined fit. \textit{From top to bottom}: Simulated galaxy at z=2.20 with a dust extinction of 0.2 and Age of 0.404 Gyr,  simulated galaxy at z=3.15, with E(B-V)=0.4 and Age=0.321 Gyr and a simulated galaxy at z=4.26, E(B-V)=0.2 and Age=0.4 Gyr. In each map, the red circle shows the maximum of probability, the 1$\sigma$ contour is shown by the blue line and the input couple (E(B-V),Age) is shown by the black star. The age we are using here corresponds to the $\mathcal{A}_{onset}$ definition. In this example, the combined fit is the only one able to reproduce the input values and also the only one where the 1$\sigma$ is the smallest.}%
  \label{E_ind_maps}%
\end{figure*}

Three representative examples are presented in Figure \ref{E_ind_maps} with the PDMs of 3 simulated galaxies at z=2.20, z=3.15 and z=4.26, with (E(B-V), Age) pairs of (0.2,0.40), (0.4,0.32) and (0.2,0.4), respectively. 
The shape of the $1\sigma$ contours are clearly the result of the age-dust degeneracy. Indeed, to keep a similar template shape, an increase of the dust extinction has to come with a lower age. 

As discussed in the previous Section,  the spectroscopy only is not able to measure the age with a good accuracy. Since, age and dust are closely linked together, the accuracy on the dust estimate is not so high either. We clearly see that the $1\sigma$ contour at low age corresponds to higher values of E(B-V) and the contours at ages higher than the simulated age correspond also to smaller dust extinction values. This is the result of the age-dust degeneracy. Moreover, over the three example presented here only one clearly shows an input pair in the $1\sigma$ contour while the two other ones are at the limit.

The presence of strong secondary peaks in the PDMs build from the fit of the photometric data alone confirms that the photometry alone is not able to resolve the age-dust degeneracy. The map based on the photometry only shows that the spectral range is large enough to be sensitive to dust extinction variations but remains a poor constraint in the blue part of the spectral domain because of the limited spectral resolution of the photometry (typically one point per $\sim$1000\AA). 

Finally, the PDMs based on the combined fit of UV rest-frame spectroscopy and photometry shows only one peak with a much smaller $1\sigma$ contour, as shown on examples presented in  Figure \ref{E_ind_maps}). Solutions with lower age and higher dust extinction or with higher dust extinction and lower ages are mostly ruled out. It is remarkable that the combination of photometry and spectroscopy data allows to break the age-dust degeneracy and to produce a pronounced probability peak allowing to recover the simulated (Age,dust) pair. Rest-frame UV spectra provide a stronger constraint than the photometry in a wavelength domain where the effect of dust extinction is strongly varying.

\begin{figure}[h!]
  \centering
  \includegraphics[width=8cm,height=8cm]{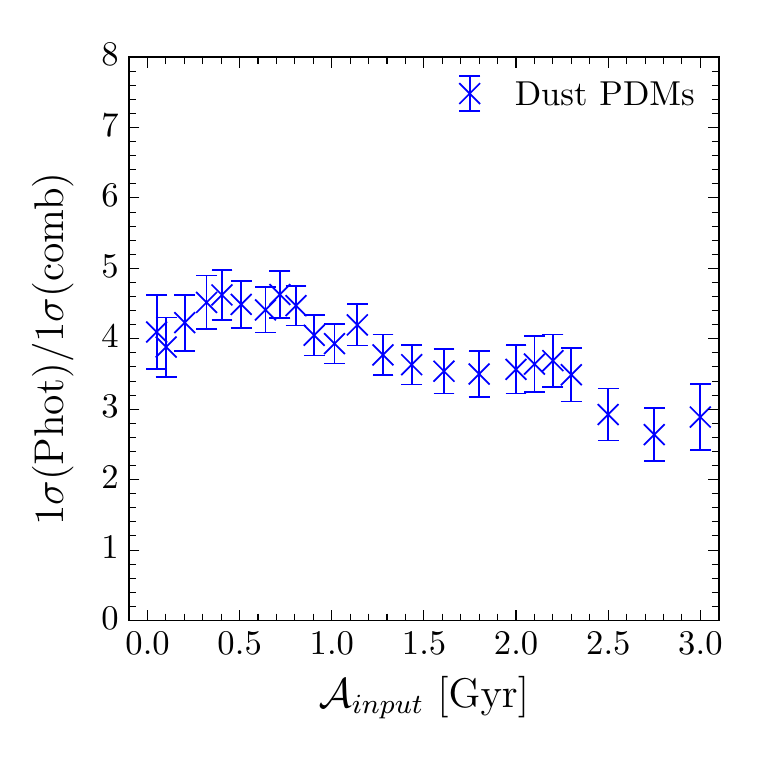}
  \caption{Comparison of the $1\sigma$ area from the fit on the photometry and the fit on the combined data as function of the input $\mathcal{A}_{onset}$ for the age-dust relation. At any age, the $1\sigma(phot)$ area based on the photometry is on average 4 times bigger than the $1\sigma(comb)$ area, based on the combined fit. The errorbars are computed from the median absolute deviation.}%
  \label{area_E}%
\end{figure}

o study the age-dust degeneracy over the whole simulated sample we turn now into the analysis of the three properties described in the previous Section. Table \ref{GloPropAE} shows two of these properties for the Age-Dust PDMs. We give the mean Npeak and the fraction of simulations in which the input couple (Age-E(B-V))$_{in}$ is contained in the 1$\sigma$ contour. We compare here only the two \textit{best} fit mode, photometry and the combined photometry plus spectroscopy datasets. 

This table shows that in the combined fit case, the $1\sigma$ contour of the PDM contain the input couple in more than 80\% of the time. This number drops below 70\% when the photometry only is used. The combined fit is therefore the fit mode that is more likely to retrieve the right parameters. Then, the median number of peaks encompassed by the 1$\sigma$ contour is of 1 for the combined fit and of 2 for the pure photometric fitting. This means that when we analyse the PDMs of the combined fit the output couple is in general the only one possible while the photometric fitting gives another possible peak due to the degeneracy.

\begin{table}[h]
\caption{Global Properties of the Age-Dust PDMs in our simulations. }
\label{GloPropAE}    
\centering                       
\begin{tabular}{ccc}
\hline 
\hline 
Property & Photometry & Combined  \\ 
\hline
\hline  
(Age-E(B-V))$_{in}$ $\in$ $1\sigma$ & 66\% & 81 \% \\
Median N$_{peak}$ & 2 & 1\\
\end{tabular} 
\end{table}

Finally, we compute the area limited by the 1$\sigma$ contours of the individual PDMs. We compare the size of the $1\sigma$ area between the fit on the photometry only and the fit on the combined data. Figure \ref{area_E} shows the ratio between the size of the 1$\sigma$ area of the PDMs on the photometry alone and on the combined data.
We find that on average, $1\sigma$ contours based on the photometry cover 3.9 times larger area than $1\sigma$ contours based on the combined fit. 

Our simulations therefore demonstrate that, at high redshifts $z>2$ when the age of the universe is significantly reduced, the combination of spectroscopy and photometry generally results in probability distributions maps with a single peak including the simulated pair of parameters. This probability peak is significantly narrower than when using the photometry only. We conclude that the Age-dust degeneracy is significantly reduced when using the combined spectroscopic and photometric data.

\subsubsection{Age-Metallicity degeneracy}
\begin{figure*}[h!]
  \centering
  \includegraphics[width=15cm,height=6cm]{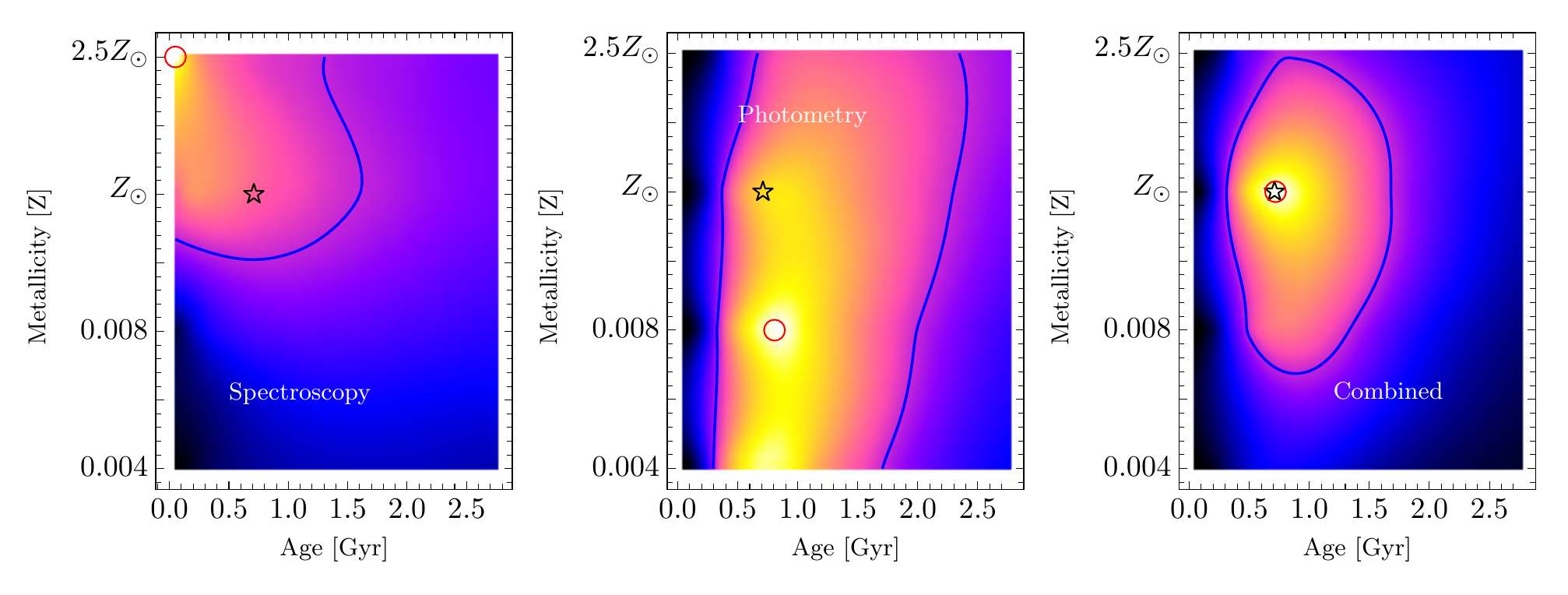}
  \includegraphics[width=15cm,height=6cm]{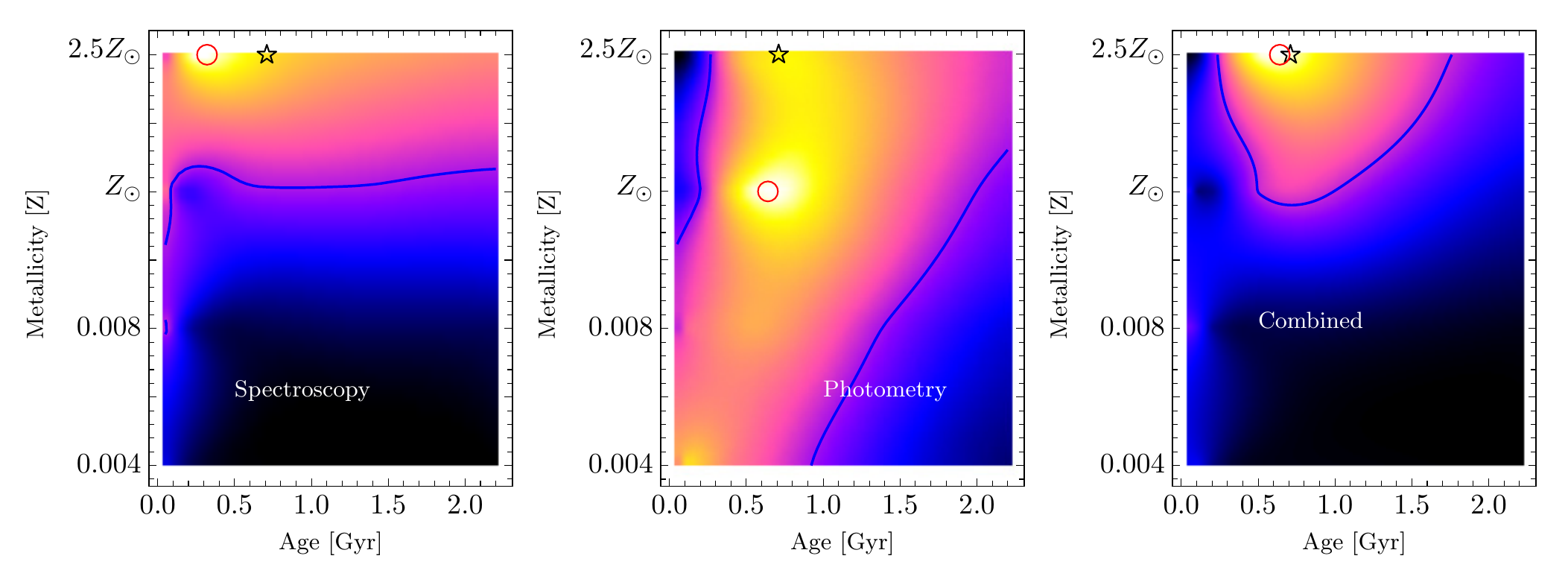}
  \includegraphics[width=15cm,height=6cm]{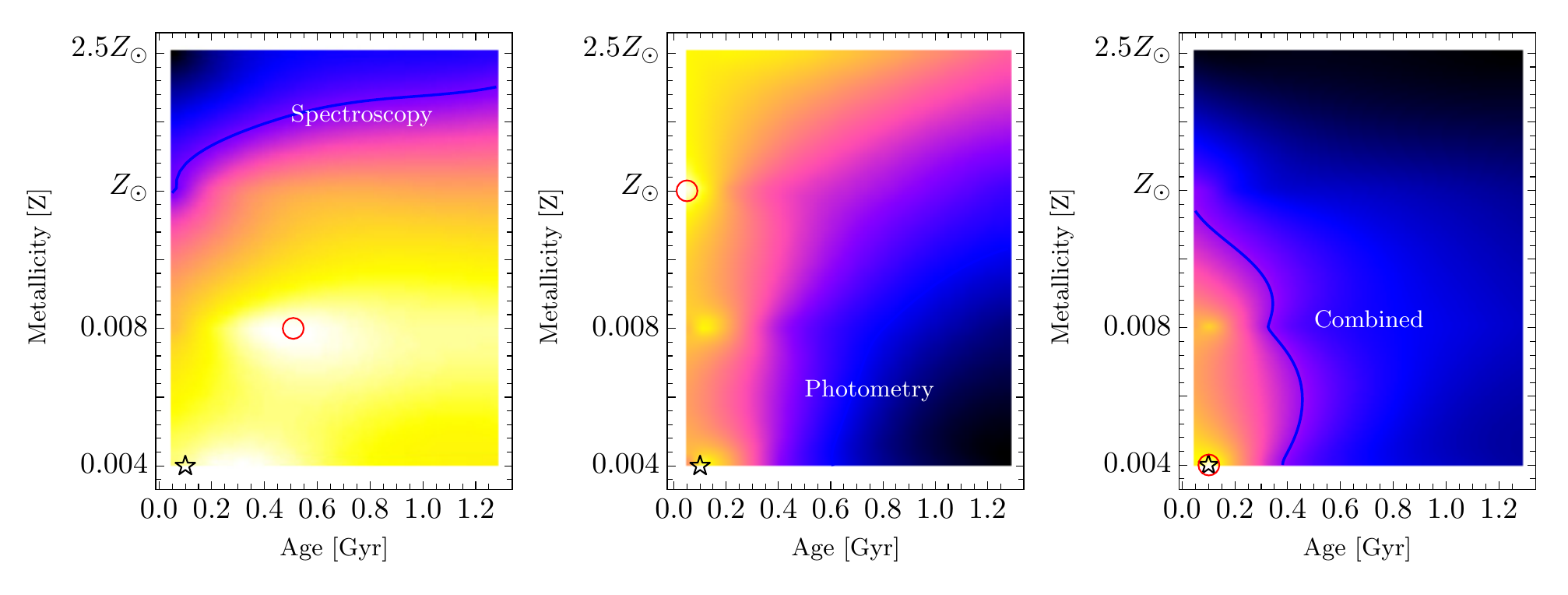}
  \caption{Same as Fig. \ref{E_ind_maps} but for the age-metallicity relation. \textit{From top to bottom}: Simulated galaxy at z=2.25 with a metallicity of Z$_{\odot}$ and an age of 0.7 Gyr, simulated galaxy at z=3.45 with a metallicity of 2.5Z$_{\odot}$ and an age of 0.7 and finally a simulated galaxy at z=4.47 with a low metallicity of Z$_{\odot}$/5 and an age of 0.1 Gyr.}%
  \label{Z_ind_maps}%
\end{figure*}

We now study the age-metallicity relation in the same way as for the age-dust degeneracy. The degeneracy between age and metallicity has long been recognized as a limitation in computing ages \citep{Worthey94}. As a stellar population in a galaxy evolves the metallicity increases steadily, therefore to reproduce the SED of a galaxy it is possible to compensate the general reddening of the SED from metallicity enhancement by a decrease of the age of the galaxy and still get a very similar SED. 
Figure \ref{Z_ind_maps} shows the individual maps for three representative simulated galaxy at z=2.25, z=3.45 and z=4.47. The age-metallicity pair of these mock galaxies are (0.7,Z$_{\odot}$), (0.7, 2.5Z$_\odot$) and (0.1, Z$_\odot$/5). The colour coding is the same as Figure \ref{E_ind_maps}. 

The PDMs resulting from the fit on the spectroscopy show that the maximum of the map is on average 0.7 Gyr away from the input value and the $1\sigma$ contour spans all the age range possible and hence the resulting parameters are poorly constrained. The spectroscopy alone shows that it is almost impossible to put a constraint on the age. The peak of the PDMs are always at a different place as the input pair and both parameter are not retrieved simultaneously.

The photometry alone does not allow for the recovery of the input parameters either. The PDMs show that ages are closer to the input, but the metallicity does not correspond to the input value. Moreover, we observe that the probability for other metallicities values remains high and hence the discrimination power is limited. While the peak of the PDM are in general close in term of age the metallicity is always different as the input one. The $1\sigma$ contours are very extended which signal poor constraints on the final parameters.

Finally the combined fit of the photometry and the spectroscopy is the only mode that correctly recovers the input parameters. The $1\sigma$ contours are well centred on the input values and narrower than the ones on the fit of the photometry only.

\begin{figure}[h!]
  \centering
  \includegraphics[width=8cm,height=8cm]{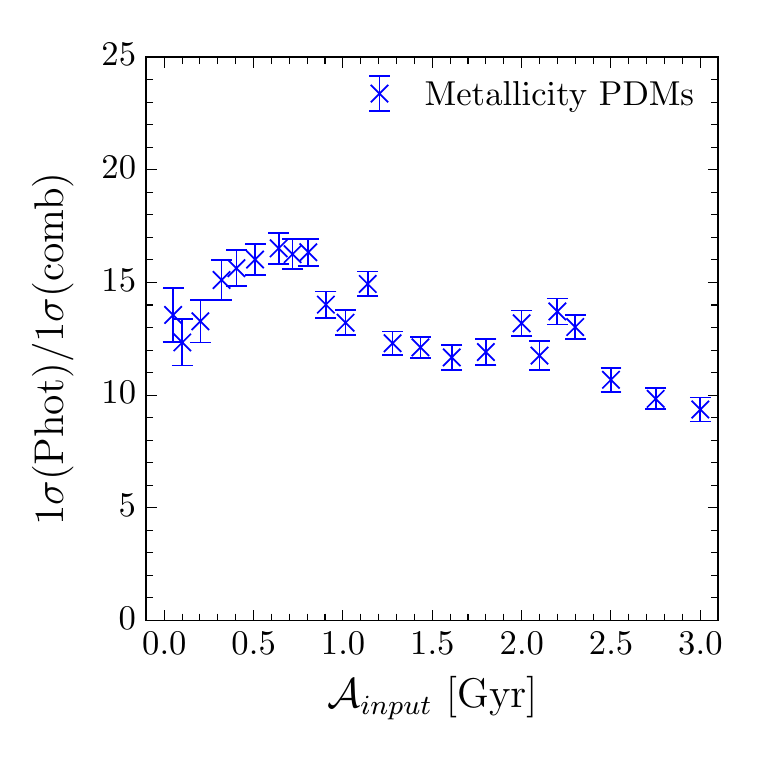}
  \caption{Comparison of the $1\sigma$ area from fit on the photometry and the fit on the combined data as function of the input $\mathcal{A}_{onset}$ for the age-metallicity relation. At any age, the $1\sigma(phot)$ area based on the photometry is on average 13.3 times bigger than the $1\sigma(comb)$ area, based on the combined fit. The error-bars are computed from the median absolute deviation.}%
  \label{area_Z}%
\end{figure}

The analysis of the age-metallicity PDMs is summarized in Table \ref{GloPropAZ} and figure \ref{area_Z}. The median number of peak in the contour produced by the combined fit is unity while it increases to 2 when fitting the photometry only. In the case of the combined fit we retrieve the input couple in the contour for 78\% of the cases, a small improvement compared to the photometry only (75\%). However the ratio between the area of the 1$\sigma$ contour from the photometry only and the combined fit is in on average $\sim13$ (Figure \ref{area_Z}). This analysis indicates that the combined fit is able to retrieve the right simulated parameter couple with a significantly better significance than when using the photometry alone.

\begin{table}[h]
\caption{Global Properties of the Age-metallicity PDMs in our simulations. }
\label{GloPropAZ}    
\centering                       
\begin{tabular}{ccc}
\hline 
\hline 
Property & Photometry & Combined  \\ 
\hline
\hline  
Median N$_{peak}$  & 2 & 1\\
(Age-Z)$_{in}$ $\in$ $1\sigma$ & 75\% & 78 \% \\
\end{tabular} 
\end{table}

It is evident but well worth saying that the accuracy of physical parameters measurements will keep improving as more and more of the spectral domain is covered with spectroscopy. The main limitation of the SED fitting method will then likely remain the systematic uncertainties linked to the simplifying assumptions used to produce model spectra representative of the true galaxy population.

\section{The age distribution of galaxies with $2 \leq z \leq 6.5$ in VUDS}
\label{age}

\subsection{Fitting process}
\label{fitproc}
We fit the 3597 VUDS galaxies with the GOSSIP+ software (Section \ref{GOSSIP+}) using the combined fit of the available broad-band photometry and the spectroscopy. We use two different population synthesis models: BC03 \citep{BC03}; these libraries are presented in Table \ref{lib_epoch}.	

\begin{table}[h!]
\caption{Libraries from population synthesis models BC03 and M05 used to estimate ages of VUDS galaxies}
\label{lib_epoch}    
\centering                       
\begin{tabular}{c c}       
\hline\hline                
Parameter & BC03 \\   
\hline                       
IMF & \multicolumn{1}{c}{Chabrier}  \\ 
Metallicity &  0.004,0.008,0.02, 0.05  \\ 
SFH & \multicolumn{1}{c}{Delayed with $\tau_{SFH}$ $\in$ [0.1;5.0] Gyr} \\
E(B$-$V) & \multicolumn{1}{c}{0.0 to 0.5}\\ 
Ages (Gyr) & \multicolumn{1}{c}{0.05 to 4.0}  \\ 
IGM & \multicolumn{1}{c}{7 models / z}  \\ 
\hline                                  
\end{tabular}
\end{table}

\begin{figure*}[h!]
\centering
\includegraphics[width=15cm,height=20cm]{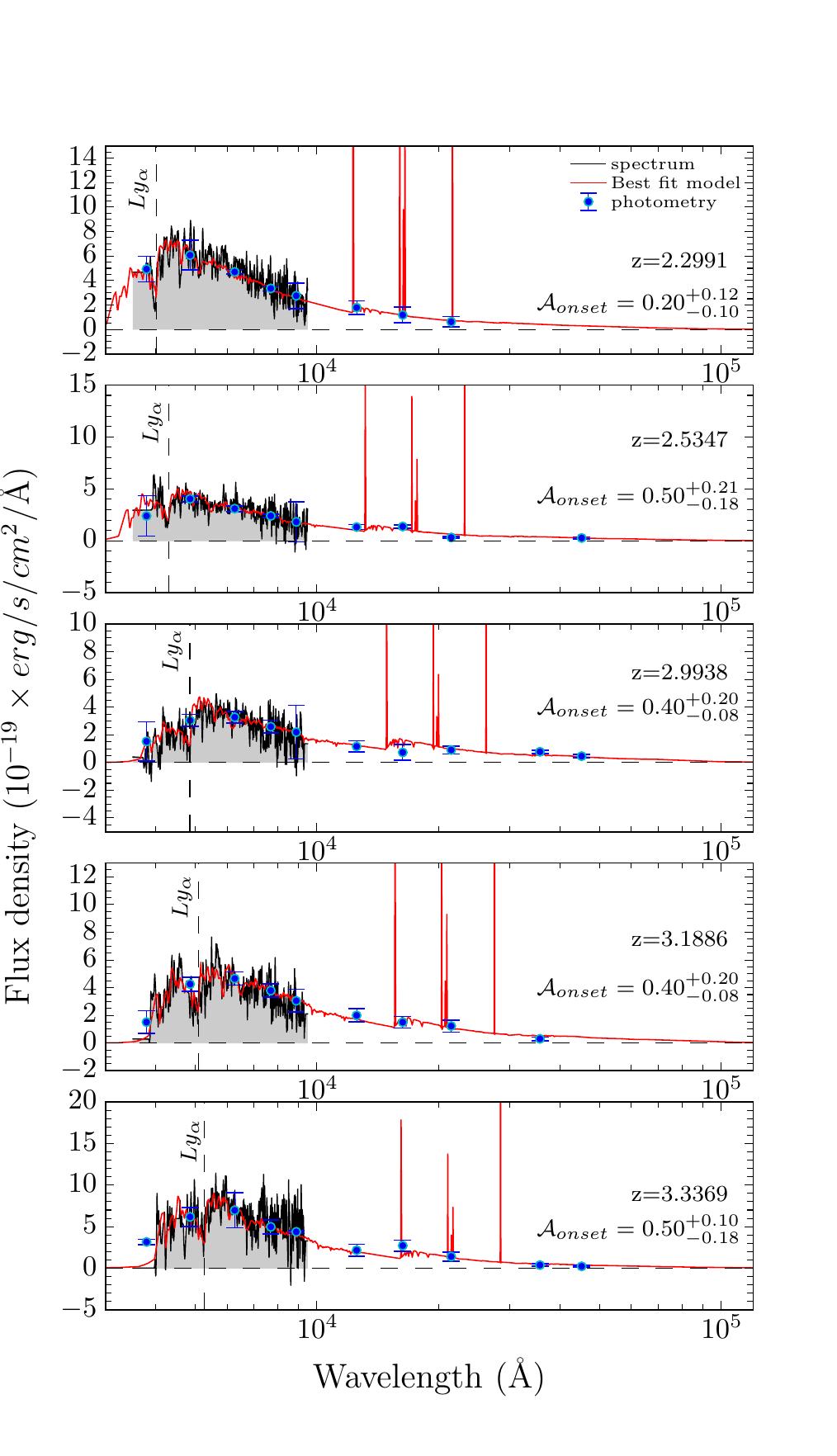}
\caption{Five example of fit in our galaxy sample between z=2.2991 and z=3.3369. In each panel the black line represents the VUDS spectroscopy, the blue points are the photometric points and the best fit template is in red. For each case, the template fit represents well both photometric and spectroscopic data.}
\label{FIT}
\end{figure*}

\begin{figure*}[h!]
\centering
\includegraphics[width=15cm,height=20cm]{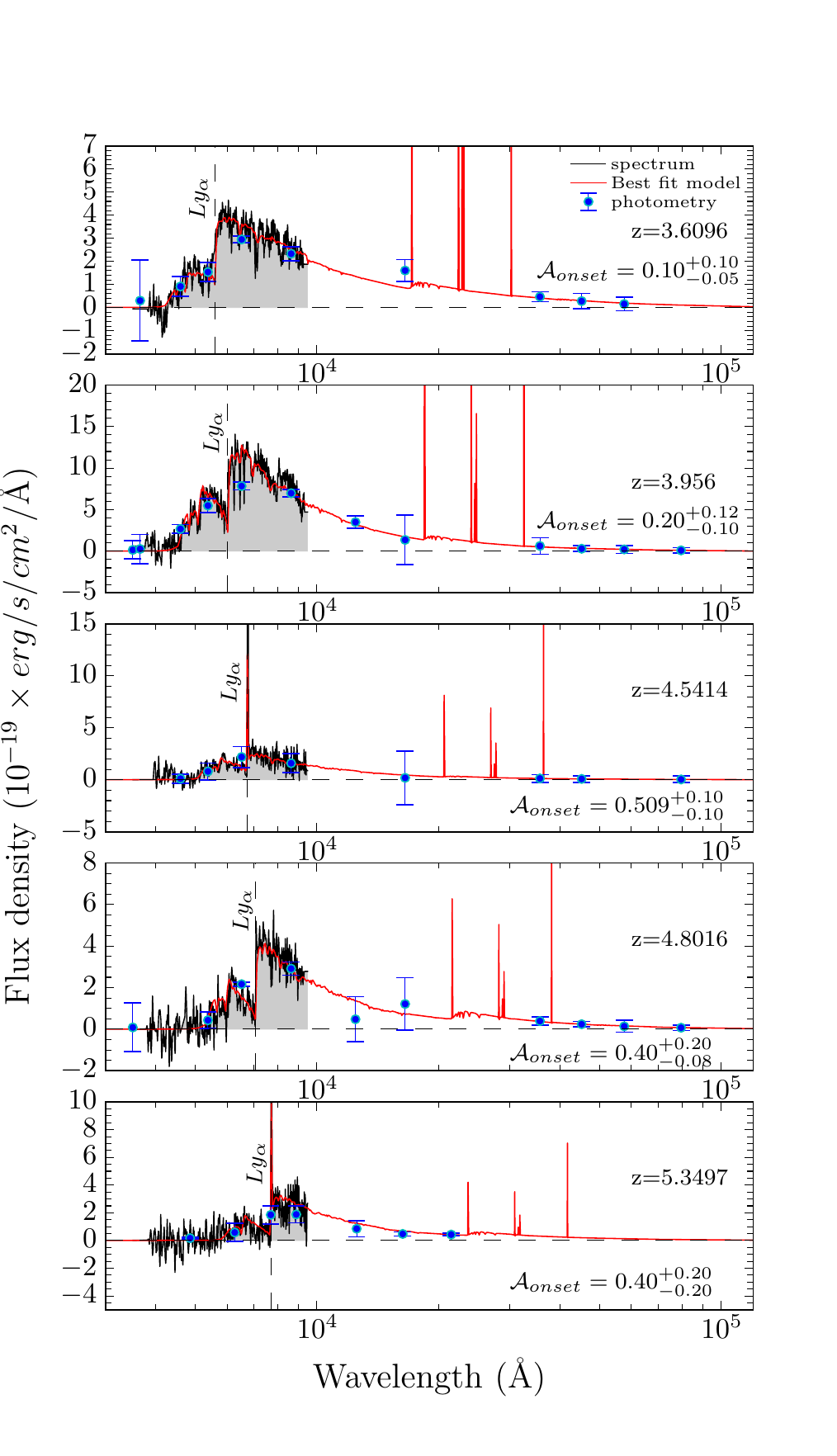}
\caption{Same as figure \ref{FIT} for galaxies between z=3.6 and z=5.34.}
\label{FIT2}
\end{figure*}

As an example of the fitting result we present the fit of 10 galaxies in our sample based on both photometry and spectroscopy in Fig.\ref{FIT} and \ref{FIT2}, covering the redshift range of this study. The agreement between the data and the best fit model is good in all cases, representative of the whole sample. 
 
\begin{figure*}[h!]
\centering
\includegraphics[width=\hsize]{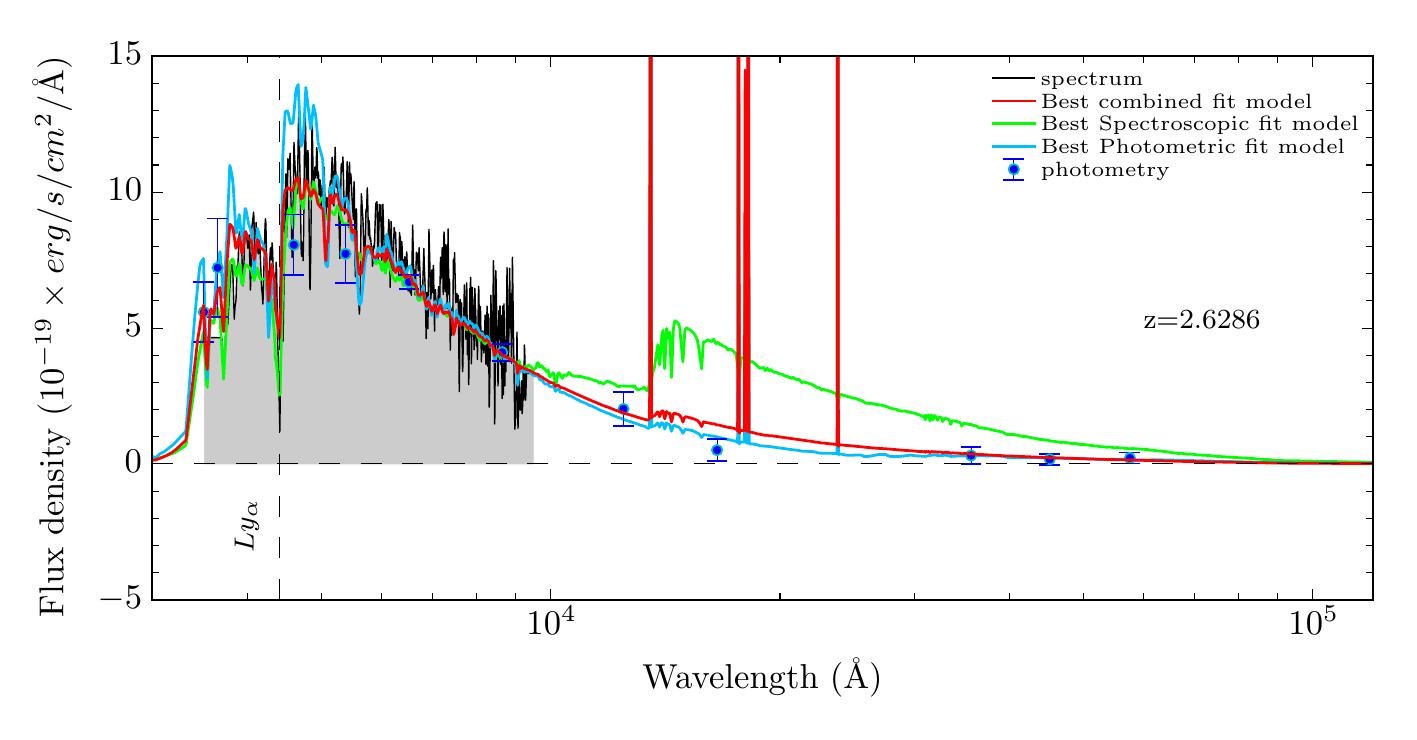}
\caption{Comparison of the Fits from the three types of fitting modes on a galaxy at z=2.6286. The black line represent the observed spectroscopy and the blue points represent the observed photometry. In light blue, red and green we show the best fit model of the fit on the photometry, combined data and spectroscopy, respectively. The fit on the combined data is the only able to reproduce both type of data.}
\label{FIT_comp}
\end{figure*}

An example of the effect of combining photometry and spectroscopy with a comparison of each type of fit (spectroscopy only, photometry only and combined data) is presented in Figure \ref{FIT_comp}. The fit on the spectroscopy data represents very well the data in the wavelength range of the spectroscopic data, but the redder part ($\lambda>9500$\AA) is not well constrained, by definition, and the best fit model is therefore far from the photometry in this region. The best fit model based on the photometric data does indeed fit the photometry very well. Nevertheless, the  wavelength range $3500<\lambda<9500$\AA ~(corresponding to the range of the spectroscopy data) is poorly constrained and the Lyman$-\alpha$ region is not well reproduced because the sampling by photometric data is quite coarse in this domain. Finally, this example shows that the fit on the combined data is the best able to reproduce both the infrared photometry and the UV-restframe spectroscopy.

\subsection{Stellar Mass and SFR}
\label{fitMSFR}
\begin{figure}[h!]
\centering
\includegraphics[width=\hsize]{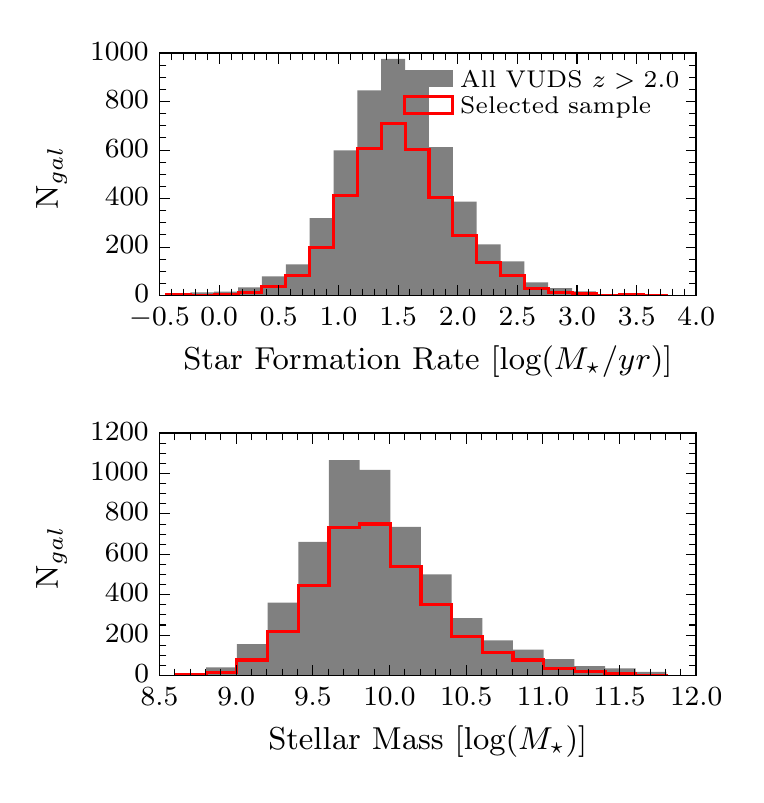}
\caption{Stellar Mass and SFR distributions for galaxies in the selected VUDS sample (red histograms), compared to the full VUDS sample (shaded grey histograms). The two distributions are similar showing that our selection is not imposing any particular bias.}
\label{mass_SFR}
\end{figure}

As the quarter-mass-age and mass-weighted-age measurements depend strongly on mass estimation, we  first present the M$_{\star}$ and SFR distributions.
The stellar mass distribution is presented in the bottom panel of Figure \ref{mass_SFR} and the star formation rate is shown in the upper panel. Stellar masses range from $\log$M$_{\star}\sim$8.3 to $\sim$11.9 with a median $\log$M$_{\star}=9.8$.
The SFR ranges from $\log$SFR$\sim-0.25$ to $\sim$3.50, with a median $\log$SFR$=1.45$. 
The M$_{\star}$ and SFR distributions in the sample used in this paper (Section \ref{galsample}) are similar to the distributions of all VUDS galaxies and hence our selection is not producing any particular population bias.

\subsection{Ages from the fit of spectra and photometry combined}

\begin{figure}
\centering
\includegraphics[width=\hsize]{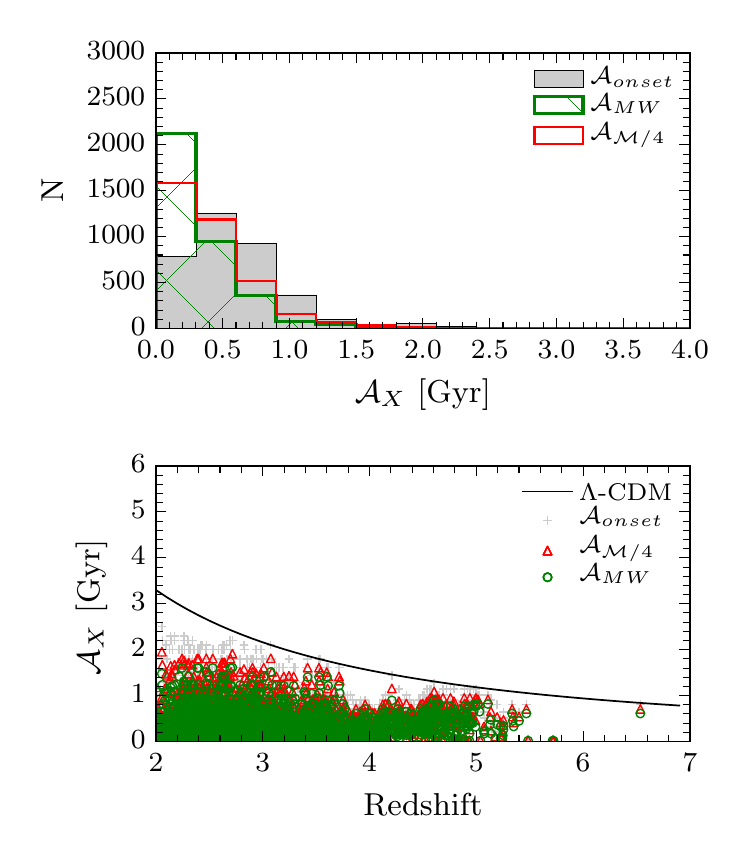}
\caption{Age distributions of 3597 VUDS galaxies computed from the combined fit of the spectroscopy and the photometry as described in Section \ref{GOSSIP+}. \textit{Top panel:} The three distributions in red, hatched green and filled grey histograms correspond to the three definitions of age we use in our paper and defined in Section \ref{sec_age_def}, $\mathcal{A}_{onset}$, $\mathcal{A}_{MW}$ and $\mathcal{A}_{\mathcal{M}/4}$, respectively. \textit{Bottom panel:} Age distributions of VUDS galaxies as a function of observed redshift $z_{obs}$. The three age definitions, $\mathcal{A}_{onset}$, $\mathcal{A}_{MW}$ and $\mathcal{A}_{\mathcal{M}/4}$ are presented in red, green and grey, respectively. The line represents the age of the Universe in the $\Lambda$-CDM cosmological model. The lowest age corresponds to the smallest possible age allowed by our model library (50 Myr). The upper limit is in general lower than the age of the Universe.}
\label{age_dis_and_zobs}
\end{figure}

The age distribution of VUDS galaxies is presented in the top panel of Figure \ref{age_dis_and_zobs} for the three definitions of age $\mathcal{A}_{onset}$, $\mathcal{A}_{MW}$ and $\mathcal{A}_{\mathcal{M}/4}$ (see Section \ref{sec_age_def}). 

For $\mathcal{A}_{onset}$, the ages range from 0.05 Gyr (the lowest age allowed in the fit) to ages higher than 2 Gyr. The former represents a very low fraction of our galaxies around $\sim$1\%. Around 10\% of galaxies have ages in the range 1 Gyr $<\mathcal{A}_{onset}<$ 2 Gyr while $\sim$89\%, are younger than 1 Gyr old. 
The $\mathcal{A}_{\mathcal{M}/4}$ ages are younger than 2 Gyr with $\sim$5\% in the range 1 Gyr $<\mathcal{A}_{onset}<$ 2 Gyr, the rest of the sample being younger than 1 Gyr.

Using the $\mathcal{A}_{MW}$ definition, $\sim$2\% of galaxies are between 1 and 2 Gyr, and most of the sample is younger than 1 Gyr old.

Differences observed between age distributions with the three different age definitions are as expected since the mass-weighted age is smaller than the quarter-mass age, the latter being smaller than the onset age.

The bottom panel of Figure \ref{age_dis_and_zobs} shows the distributions of age as a function of redshift. Ages continuously cover the range from the lowest ages allowed in our fit (0.05 Gyr) to ages up to 2 Gyr at $z\sim2-3$, the lowest redshift end of our sample. We do not observe galaxies right at the age corresponding to the age of the Universe in $\Lambda$CDM. This indicates that our data and measurement process provide enough age resolution and accuracy to avoid a saturation effect on the largest ages; less than 10 objects have ages close to the age of Universe at the observed redshift.


\section{The formation redshift function (FzF)}
\label{sec_fzf}

\subsection{Determining the redshift of formation from galaxy ages}

\begin{figure}[h!]
\centering
\includegraphics[width=\hsize]{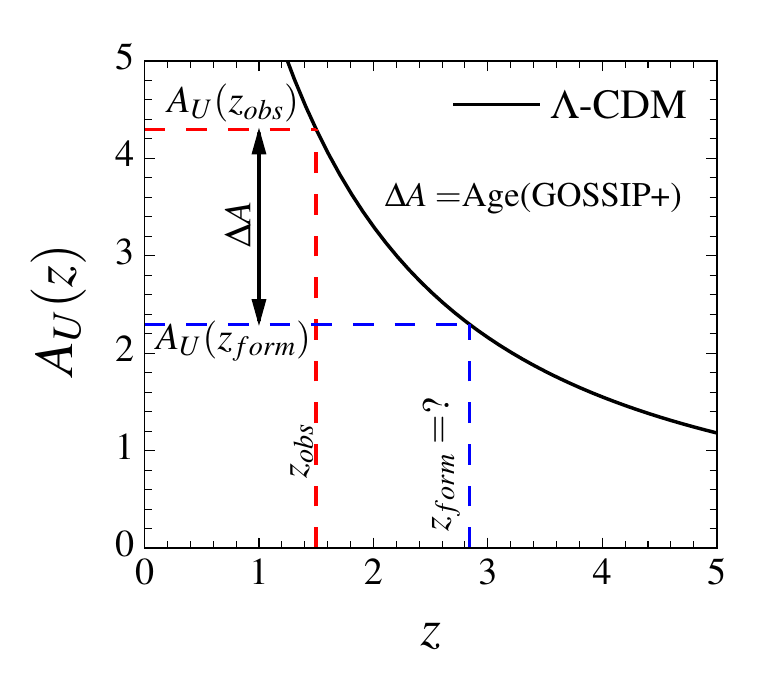}
\caption{Method used to compute the formation redshift. The observed redshift is $z_{obs}$ and the age of the Universe at this redshift is given by A$_{U}(z_{obs})$. The age of the galaxy measured by GOSSIP+ is Age(GOSSIP+). A$_{U}(z_{f})$ is the age of the Universe at $z_{f}$.  The black curve represents the age of the Universe as a function of redshift in the $\Lambda$-CDM model. From the measurement of $z_{obs}$ and the age given by GOSSIP+ we can infer the formation redshift by successive iterations (see text).}
\label{zform_met}
\end{figure}

From the age distributions presented in Section \ref{age} we now compute the formation redshift $z_f$ of each galaxy, combining the age of the galaxy and the observed redshift $z_{obs}$. The way $z_{f}$ is measured is illustrated in Figure \ref{zform_met} where we link the redshift at which the galaxy is observed, $z_{obs}$, the age of the galaxy, and the formation redshift $z_{f}$. The formation redshift is calculated by substracting the age of the galaxy (determined with GOSSIP+, $\Delta A$ in the Figure), from the age of the Universe A$_{U}(z_{obs})$ at the observed redshift $z_{obs}$. The redshift corresponding to the resulting age, A$_{U}(z_{f})$, is the formation redshift.

The computation of the formation redshift is quite sensitive to age errors. 
The uncertainty on $z_f$ is defined as:
\begin{equation}
\Delta z_f= z_f(Age+\delta Age)- z_f(Age)
\end{equation}
$\Delta z_f $ ranges from $\sim0.1$ at $z\sim2$ to $\sim 2$ at $z\sim6$ for an uncertainty in age measurement of $\delta Age=0.1$Gyr, and this more than doubles for $\delta Age=0.3$Gyr. A small change of age at high redshift may therefore significantly change the formation redshift estimation. However, the relative distribution of $z_f$ in a star-forming galaxy population should not be affected, as discussed below.

\begin{figure}[h!]
\centering
\includegraphics[width=\hsize]{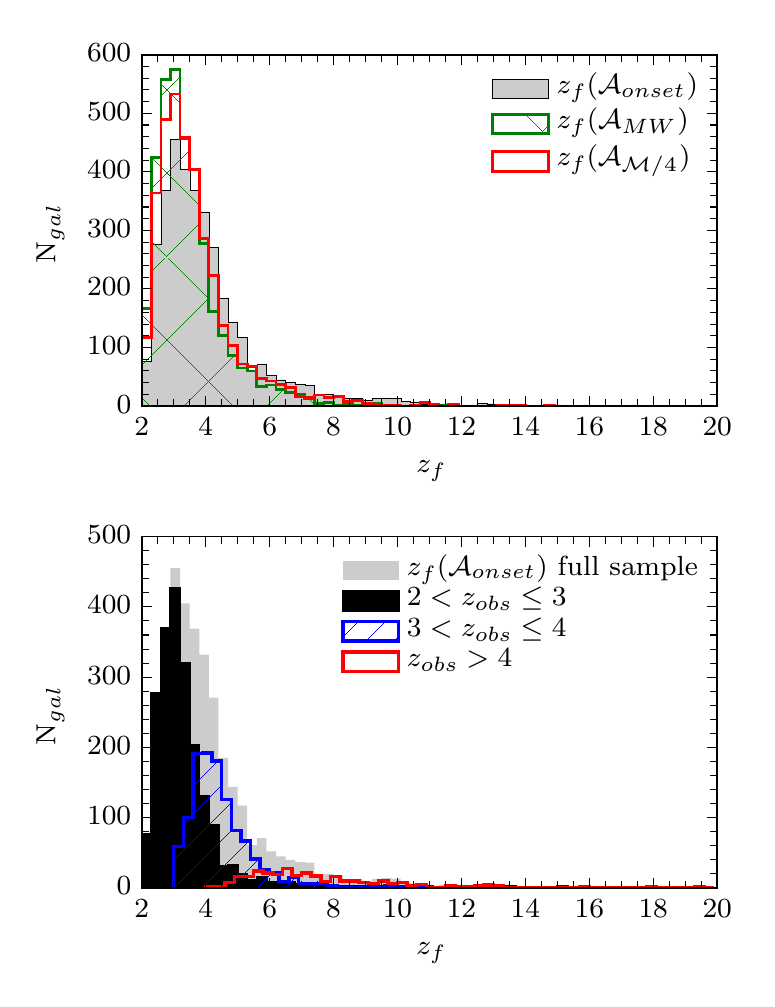}
\caption{Formation redshifts of the VUDS sample.\textit{Top panel}: Formation redshift distributions from the three age definitions $\mathcal{A}_{onset}$, $\mathcal{A}_{MW}$ and $\mathcal{A}_{\mathcal{M}/4}$ (grey filled, hatched green, and red empty histograms, respectively). 
\textit{Bottom panel}: Comparison of the formation redshift distribution (from $\mathcal{A}_{onset}$) in three \textit{observed} redshift bins: $2<z_{obs}\leq3 $, $3<z_{obs}\leq 4 $ and $z_{obs}>4$, shown by the filled black histograms, blue hatched histogram and red empty histogram
}
\label{zf_data}
\end{figure}

We apply this method to compute $z_f$ for the 3597 galaxies in our VUDS sample and obtain the distribution of formation redshifts presented in Figure \ref{zf_data}. 
In the top panel the maximum formation redshift defined from $\mathcal{A}_{onset}$ is $z_f=19.5$ and a significant rise in the number of galaxies occurs at $z_f<10$. The median of the distribution is $<z_f(\mathcal{A}_{onset})>\sim 3.7$. With the $\mathcal{A}_{\mathcal{M}/4}$ definition, the highest formation redshift is at $z_f=14.8$, the median formation redshift is $<z_f(\mathcal{A}_{\mathcal{M}/4})>\sim3.36$ and the histograms starts to increase significantly at $z_f < 9$. Using $\mathcal{A}_{MW}$, the highest formation redshift is at $z_f=11.9$, the median formation redshift is $<z_f(\mathcal{A}_{\mathcal{MW}})>\sim3.22$ with a significant fraction of galaxies with $z_f\sim8$.

In the bottom panel of Figure \ref{zf_data} we present the distribution of formation redshifts for galaxies in 3 \textit{observed} redshift bins, $2<z_{obs}\leq3 $, $3<z_{obs}\leq 4 $ and $z_{obs}>4$. We observe that the formation redshift distribution shifts to higher redshifts for galaxy observed at increasing redshifts. This is easily understood when applying an increasing lower redshift bound to a continuous age distribution when going to higher redshifts, cutting out the galaxies which form at the lower redshifts. 

The observed $z_f$ distributions reported in Figure \ref{zf_data} are shaped by the redshift distribution of the VUDS sample \citep{lefevre15} produced by the survey selection function, and therefore do not represent directly the volume-average distribution of formation redshifts of the underlying population. This is derived in the next Section. 

\subsection{The distribution of formation redshifts: the formation redshift function (FzF)}

\subsubsection{The formation redshift function (FzF)}

To investigate when galaxies preferentially form we introduce a new statistical description of a galaxy population: the formation redshift function (FzF). The FzF describes the number of galaxies formed per unit volume at a formation redshift $z_f$. From the definition of $z_f$ the FzF indicates how many galaxies started forming their stars at a given $z_f$. 

The observed $z_f$ distribution needs to be corrected for the selection function of the VUDS sample applying the target sampling rate (TSR) and spectroscopic redshift success rate \citep[SSR; for a definition of TSR and SSR see][]{Cucciati2012,lefevre15} to the observed data to recover volume quantities (Tasca et al. in prep.). From the formation redshifts $z_f$ determined in the previous Section we compute the number of galaxies formed at $z_f$ in a formation redshift bin $\Delta z$ using the V$_{max}$ formalism \citep{Schmidt1968}, computing for each galaxy V$_{max}$ as the volume that the galaxy could lie in without dropping outside the survey selection limits. 

To approximately follow the evolution of the same galaxy population with redshift and keep the highest number of galaxies in the FzF estimation , we select galaxies that are more massive than  $\log_{10}$M$_{\star} \geq 9.3 \mathrm{M_{\sun}}$ at z$\sim$5. We then evolve this mass limit with redshift using the evolution of the characteristic mass of the stellar mass function as derived by \cite{Ilbert2013} and the evolution of the sSFR with redshift in the VUDS survey \citep{Tasca2014}. This leads to the following mass treshold:

\begin{equation}
\log_{10}M_{\star}>10.30-z\times0.2.
\end{equation}

\begin{figure}[h!] 
\centering 
\includegraphics[width=\hsize]{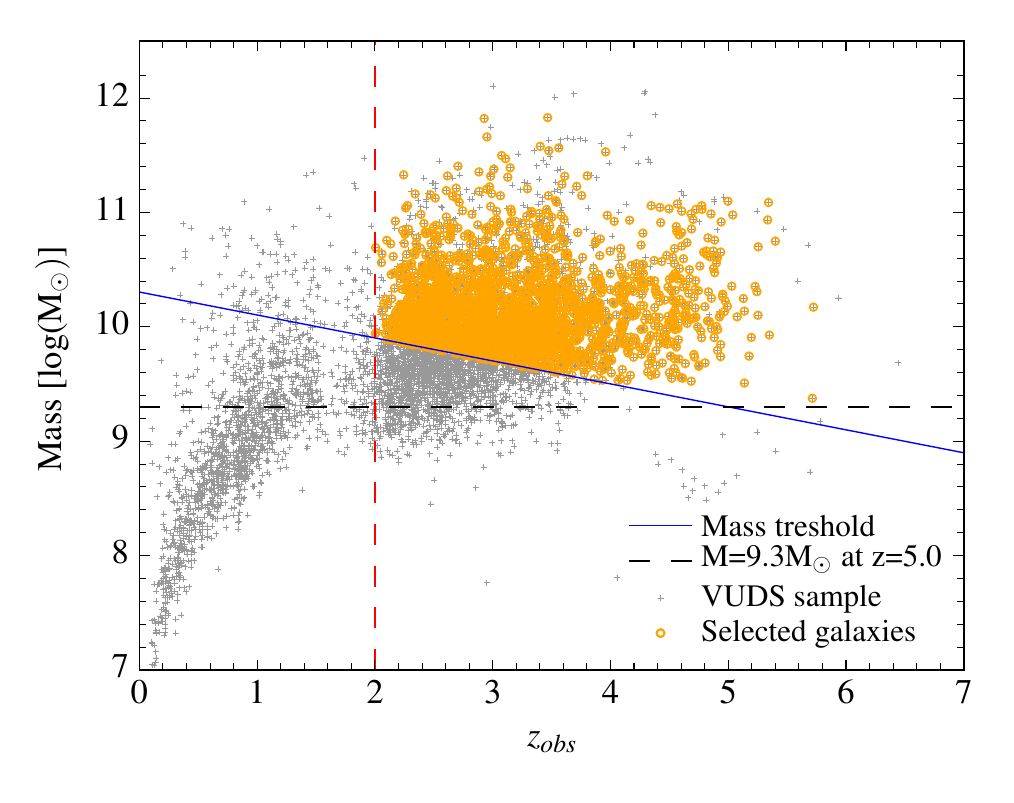}
\caption{Mass cut-off used to select the galaxy population for the construction of the formation redshift function (FzF). The grey crosses represent the full VUDS sample while the orange circles are the  galaxies selected for the computation of the FzF. These galaxies are selected to be above the blue line. The blue line represents the evolution of a galaxy with $\log$M$_{\star}=9.3$ at $z=5$ with redshift, following the evolution of the characteristic mass of the stellar mass function as derived by \cite{Ilbert2013}.}
\label{mass_cut}
\end{figure}

The selection of galaxies is shown in Figure \ref{mass_cut}.

\begin{figure*}[h!]
\centering
\includegraphics[width=\hsize]{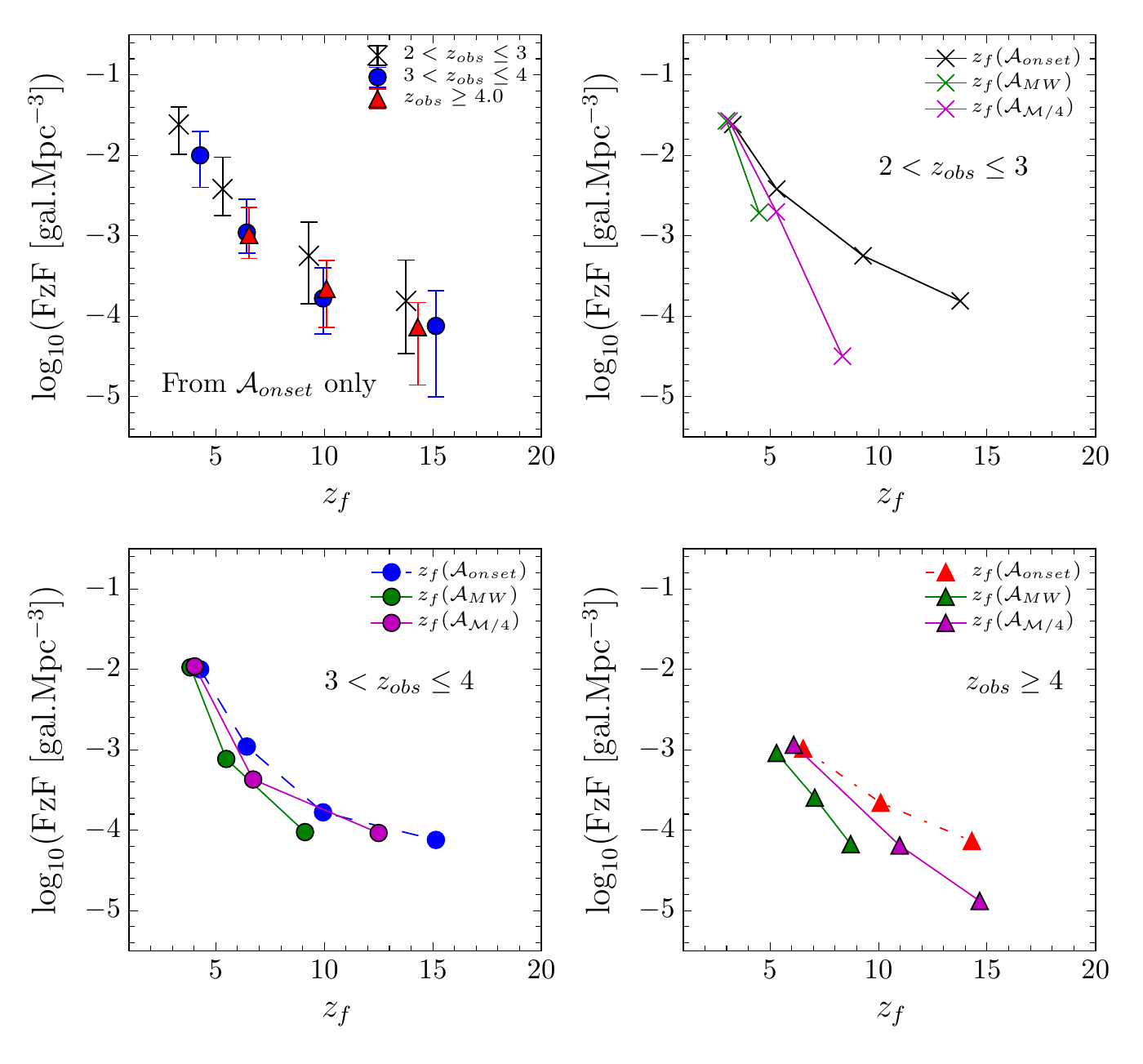}
\caption{Formation redshift functions (FzF). {\it Top left}: Formation redshift functions computed from the galaxies in 3 observed redshift bins: $2<z_{obs}\leq3$ (black crosses), $3<z_{obs}\leq4$ (blue points) and $z_{obs}\geq4.0$ (red triangles). 
The three other panels correspond to the comparison of the FzFs corresponding to the 3 different age definitions used in this paper, in increasing redshift bins. From top right to bottom right we compare the FzFs in the following observed redshift bins: $2<z\leq3$, $3<z\leq4$ and $z\geq4.0$.}
\label{FzF}
\end{figure*}

Applying this mass cut-off we selected $\sim2350$ VUDS galaxies to compute the FzF in three observed redshift ranges, $2<z\leq3$, $3<z\leq4$ and $z\geq4.0$, as shown in Figure \ref{FzF}. We find that the FzF has a very similar shape in the three observed redshift ranges, with a continuously rising number of forming galaxies from high $z_f$ down to the lower redshift considered in this study (top left panel). From redshift 10 to redshift 4 there are $\sim60$  times more galaxies forming that will later develop into galaxies with M$_{\star}>10.30-z\times 0.2$. The number of galaxies starting forming their stars at $z\sim10$ is already substantial with a value at $\sim2.88\times10^{-4}$ galaxies$\times$Mpc$^{-3}$ while at $z\sim3$ the number of forming galaxies is $\sim1.4\times10^{-2}$ galaxies$\times$Mpc$^{-3}$. 

The comparison of the FzF produced from different age definitions is presented in the 3 other panels of Figure \ref{FzF}. Different age definitions do not produce a significant change in the two highest redshift bins. While in the lowest redshift bins, we see some difference between the FzF given by $\mathcal{A}_{onset}$ on one hand and $\mathcal{A}_{\mathcal{M}/4}$ or $\mathcal{A}_{MW}$. This is in line with the formation redshift distributions presented in the previous Section.

\begin{figure}[h!]
\centering
\includegraphics[width=\hsize]{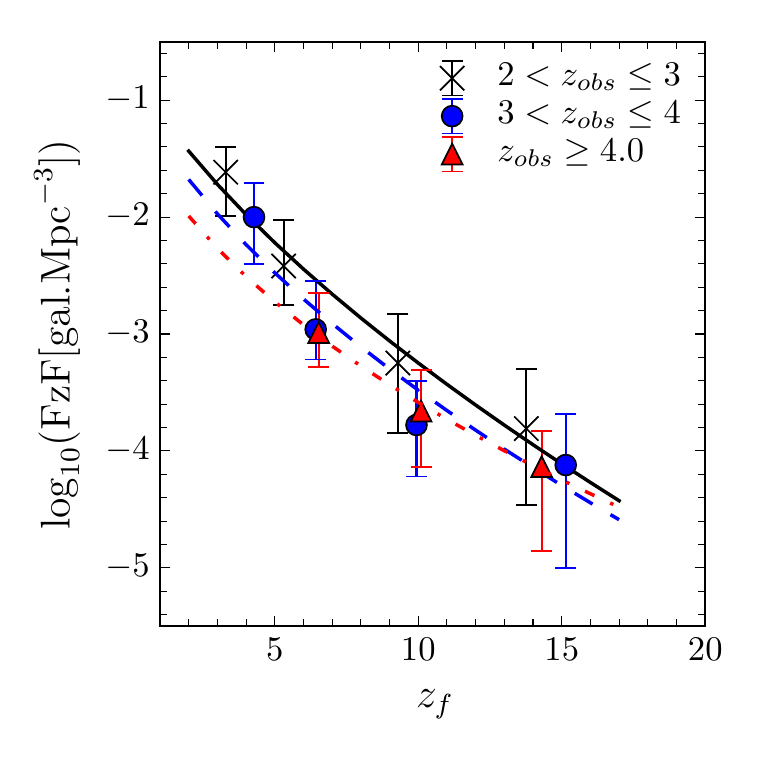}
\caption{Fit of the formation redshift functions in three observed redshift bins using $\log_{10}(f)=\alpha(1+z)^\zeta$.}
\label{FzF_fit}
\end{figure}

We parametrize the FzF by a power law $\log_{10}$FzF$=\alpha (1+z)^{\zeta}$. The results of the fit for the FzF derived from $\mathcal{A}_{onset}$ are shown in Figure \ref{FzF_fit} and we list the values of $\alpha$ and $\zeta$ for different observed redshift bins in Table \ref{alphazeta}.

\begin{table}[h!]
\caption{Parameters of the formation redshift function (FzF) fit in different observed redshift bins. The fit function is parametrized as $\log_{10}$FZF$(z) =\alpha (1+z)^{\zeta}$.}
\label{alphazeta}    
\centering                       
\begin{tabular}{c c c c c c}       
\hline\hline          
$z_{obs}$ &$<z_{obs}>$ & $\alpha$ &  $\delta\alpha$ & $\zeta$ & $\delta\zeta$\\   
\hline
\hline
$2.0<z<3.0$ & 2.63 & -0.71 & 19\% & 0.62 & 12\%\\
$3.0<z<4.0$ & 3.40 & -0.90 & 28\% & 0.56 & 23\%\\
$z>4.0$     & 4.48 & -1.21 & 10\% & 0.45 & 8\%\\
Full sample &  -   & -0.84 & 14\% & 0.58 & 10\%    \\
\hline     
\end{tabular}
\end{table}
The parameters coming from the fit of the FzF are very close to each other from one observed redshift bin to another. This strong similarity of the FzF independently of the observed redshift seems to indicate that the FzF is a universal function for this population of star-forming galaxies selected to have  $\log_{10}$M$_{\star}>10.30-z\times0.2$.

\subsubsection{Influence of the SFH and Stellar population models on the FzF}

\begin{figure}[h!]
\centering
\includegraphics[width=\hsize]{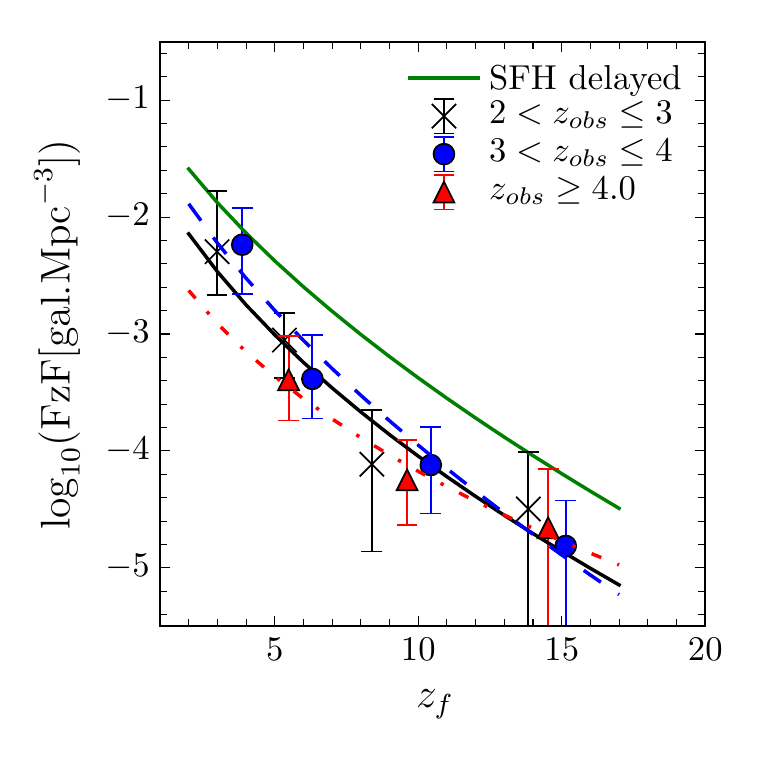}
\caption{Comparison of the FzFs from different SFHs. In green, a global FzF computed from all the point presented in Figure \ref{FzF_fit}, derived with exponentially delayed SFH. In black, blue and red, we show the FzFs derived from exponentially delayed SFH as well as their fits in 3 observed redshift bins: $2<z_{obs}\leq3$, $3<z_{obs}\leq4$  and $z_{obs}\geq4.0$, respectively.}
\label{FzF_dec}
\end{figure}

The FzFs presented above have been computed from the combined fit using exponentially delayed SFH. This kind of galaxy SFHs present both rising and declining parts. Nevertheless it is possible to study other kind of SFHs like the purely exponentially declining SFHs. We therefore reran GOSSIP+ using only exponentially declining SFHs. The resulting FzF (and associated fits) are presented in Figure \ref{FzF_dec} and Table \ref{alphazetadec}. This figure presents the comparison of the SFH from the exponentially delayed SFHs and the exponentially declining SFHs.
The new FzFs are on average 0.7 dex below the ones computed from SFH exponentially delayed. This behavior is expected. Due to the presence of only a declining part in the SFH, the ages from this type of SFH are artificially lowered. This leads to smaller formation redshifts. Consequently, at a given formation redshift in the FzF, the number of created galaxies is lower when using exponentially declining. 

\begin{table}[h!]
\caption{Parameters of the formation redshift function (FzF) fit in different observed redshift bins for the FzFs derived from exponentially declining SFH. The fit function is parametrized as $\log_{10}$FZF$(z) =\alpha (1+z)^{\zeta}$.}
\label{alphazetadec}    
\centering                       
\begin{tabular}{c c c c c c}       
\hline\hline          
$z_{obs}$ &$<z_{obs}>$ & $\alpha$ &  $\delta\alpha$ & $\zeta$ & $\delta\zeta$\\   
\hline
\hline
$2.0<z<3.0$ & 2.98 & -1.25 & 22\% & 0.48 & 18\%\\
$3.0<z<4.0$ & 3.40 & -1.00 & 21\% & 0.56 & 14\%\\
$z>4.0$     & 4.48 & -1.75 & 15\% & 0.35 & 17\%\\
\hline     
\end{tabular}
\end{table}

\begin{figure}[h!]
\centering
\includegraphics[width=\hsize]{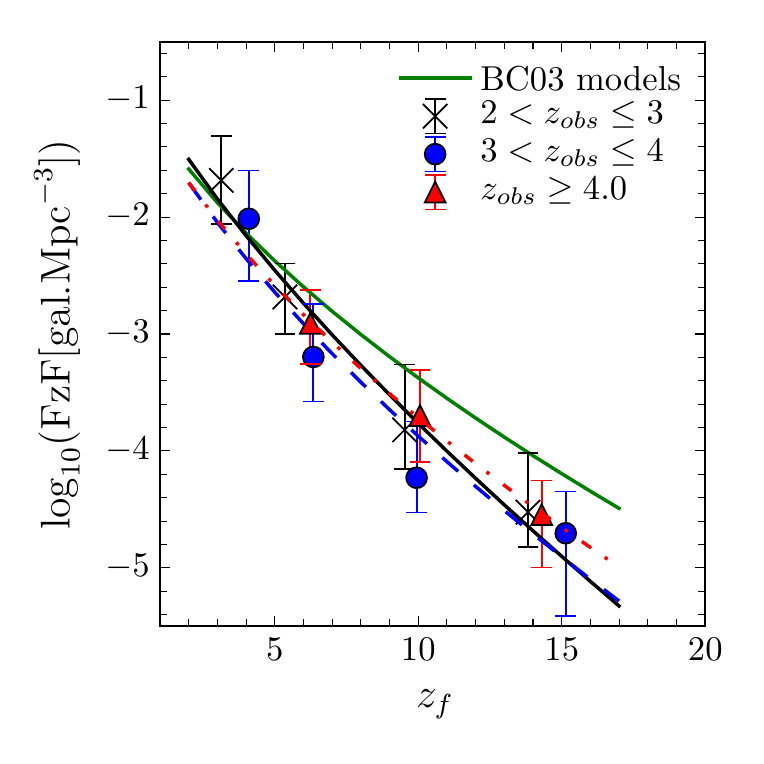}
\caption{Comparison of the FzFs from different stellar population synthesis models. In green, a global FzF computed from all the point presented in Figure \ref{FzF_fit}, derived with exponentially delayed SFH and BC03 models. In black, blue and red, we show the FzFs derived from M05 models as well as their fits in 3 observed redshift bins: $2<z_{obs}\leq3$, $3<z_{obs}\leq4$  and $z_{obs}\geq4.0$, respectively.}
\label{FzF_M05}
\end{figure}

The impact of a change of stellar population models is presented in Figure \ref{FzF_M05} and Table \ref{alphazetaM05}  using  \cite{Maraston05} models instead of BC03 models (the age we are using here are based on the $\mathcal{A}_{onset}$ definition). With these models we use the same parameters set as for BC03 models but with slightly different metallicities of 0.001,0.01, 0.02 (solar) and 0.04. 

\begin{table}[h!]
\caption{Parameters of the formation redshift function (FzF) fit in different observed redshift bins for the FzFs derived from M05 models. The fit function is parametrized as $\log_{10}$FZF$(z) =\alpha (1+z)^{\zeta}$.}
\label{alphazetaM05}    
\centering                       
\begin{tabular}{c c c c c c}       
\hline\hline          
$z_{obs}$ &$<z_{obs}>$ & $\alpha$ &  $\delta\alpha$ & $\zeta$ & $\delta\zeta$\\   
\hline
\hline
$2.0<z<3.0$ & 2.98 & -0.69 & 18\% & 0.70 & 11\%\\
$3.0<z<4.0$ & 3.40 & -0.85 & 25\% & 0.63 & 22\%\\
$z>4.0$     & 4.48 & -0.88 & 5\% & 0.60 & 4\%\\
\hline     
\end{tabular}
\end{table}

At low redshift using the M05 models leads to more forming galaxies than with the BC03 models. At high redshift, we observe the opposite: the M05 models produce less forming galaxies than the BC03 models. This change is expected as the M05 models tend to give lower ages than BC03 models (See Section \ref{simul}). Consequently the number of galaxies with high formation redshifts should be lower in the case of M05, as observed.

\section{Discussion and Conclusions: The epoch of formation of bright massive star-forming galaxies}
\label{discuss}

The distribution of ages of distant galaxies is an important piece of information in the galaxy formation puzzle. 

Our study demonstrates that the computation of galaxy ages at $z>2$ can be impressively improved when performing SED fitting on combined photometry and spectroscopy data, given the knowledge of the age of the Universe.
Performing fitting on both spectroscopy and photometry together with the knowledge of the age limit imposed by the age of the Universe at the observed redshift considerably restricts possible age models, and leads to reliable age measurements. From the analysis of PDMs, we find that degeneracies present at low redshift between age, metallicity and dust extinction  tend to be considerably reduced at high redshifts. 

We emphasize the benefit of combining UV rest-frame spectra and photometry when performing the SED fitting. While the redder photometric bands help recover the stellar mass M$_{\star}$, the detailed information available from UV spectra significantly improves the measurement of the SFR and dust extinction. 
We conclude from our analysis and simulations that the age of a galaxy at $z>2$ can be computed with a typical uncertainty of 10\% provided that important assumptions on e.g. the shape of the SFH are valid. We find that age measurements using SED fitting at these redshifts are of an accuracy comparable to that of M$_{\star}$ and SFR measurements and therefore can be taken into full consideration in our quest to understand galaxy evolution.

Using this formalism,  we use rest-frame UV spectra  from the VUDS spectroscopic survey in combination with extensive deep photometry in the COSMOS, ECDFS, and VVDS02h fields to compute ages of $\sim$3600 galaxies with $2<z_{obs}<6.5$ to study the epoch of galaxy formation. The large VUDS sample with a broad selection function allows to probe a large range of ages. At any observed redshift we find ages ranging from the youngest possible allowed by our model library at 50 Myr to the oldest allowed by the age of the Universe at that redshift. 

Assuming that the age of the Universe is known from the current best cosmological model, we then derive formation redshifts $z_f$. We explore the impact on $z_f$ of different age definitions including the age defined by the onset of star-formation, the mass-weighted-age and the quarter-mass-age. We find that these different definitions do not change the general $z_f$ distribution. 

It is striking to observe that massive galaxies with $\log_{10}$M$_{\star} \geq 10.30 - z \times 0.2$ at $z>2$ present a continuous distribution of formation redshifts (Figure \ref{zf_data}). This is an indication that galaxies can start forming their stars at any redshift and that there is no preferred epoch of galaxy formation. This is clearly evident when computing the formation redshift function FzF representing the number of galaxies per unit volume which start forming their stars as a function of formation redshift (Figure \ref{FzF}). The FzF is continuously rising as redshift decreases, well represented by a power-law $\log_{10}$FzF($z_f)=\alpha(1+z_f)^{\zeta}$, with $\zeta$ in the range from -0.61 to -1.06 at $z\sim2.6$ to $z\sim4.5$, and no preferred formation redshift can be identified. 

While galaxies appear to form at any redshifts there are a lot more galaxies forming at $z\sim3$ than at $z\sim10$.  We find that the number of forming galaxies continuously increases from $\sim2.88\times10^{-4}$ galaxies.Mpc$^{-3}$ at $z=10$, to $\sim1.4\times10^{-2}$ galaxies.Mpc$^{-3}$ at $z \sim 3$, an increase of almost 2 dex. The SFRD is directly related to the number of newly formed galaxies in the Universe which is expressed by the FzF. Assuming a constant average SFR for galaxies at the same stage of evolution across the redshift $2<z<6$, the global comoving SFRD must be proportional to the number of forming galaxies traced by the FzF, and we therefore infer from the FzF that the SFRD should also increase by $\sim$2 dex from $z \sim 10$ to $z \sim 3$. This qualitative estimate of the SFRD evolution from the FzF is in excellent agreement with the evolution of the SFRD over the same redshift range  of $\sim 1.5-2$dex as measured e.g. from the UV luminosity density \citep{Madau:14,bouwens15}. Moreover, the ratio between the SFRD and the FzF gives a SFR at $z\sim 4-6$ of $\sim8-17$ M$_{\odot}/yr$ (up to 20 M$_{\odot}/yr$ from the mass-weighted age FzF). This value is of the same order as the SFR of galaxies measured in our VUDS sample \citep[][ this paper]{Tasca2014}.

With this first detailed and systematic exploration of galaxy ages in a representative sample of star-forming galaxies in the redshift range $2<z<6.5$ we have demonstrated that the age distribution of galaxies provides important clues on the formation of galaxies in the first few billion years of evolution after the Big Bang. Reliable age estimates derived from the method proposed in this paper further expands on our tool box to characterize galaxies at any epoch towards a better understanding of galaxy formation and evolution at early cosmic times. 

Adding rest-frame optical spectra to UV rest-frame spectra, combined to broad-band photometry, will be an important part of next generation high-redshift surveys 
as this will further improve the accuracy of parameters from SED fitting. The accuracy of physical parameter measurements such as Age, M$_{\star}$, SFR or dust content will keep improving as wider spectroscopic coverage become available from NIR spectrographs on ground-based telescopes or with the James Webb Space Telescope. This will be an important contribution to further improve our understanding of galaxy formation.



\begin{acknowledgements}
We thank Claudia Maraston and Roser Pell\`o for useful discussions.
This work is supported by funding from the European Research Council Advanced Grant ERC--2010--AdG--268107--EARLY and by INAF Grants PRIN 2010, PRIN 2012 and PICS 2013. AC, OC, MT and VS acknowledge the grant MIUR PRIN 2010--2011.  
This work is based on data products made available at the CESAM data center, Laboratoire d'Astrophysique de Marseille, France. 
\end{acknowledgements}

\bibliographystyle{aa}
\bibliography{epochOLF}

\end{document}